\documentclass[final]{agujournal2019}
\usepackage{url} %this package should fix any errors with URLs in refs.
\usepackage{lineno}
\usepackage{amsmath,amsfonts,amssymb}
\usepackage[inline]{trackchanges} %for better track changes. finalnew option will compile document with changes incorporated.
\usepackage{soul}
\usepackage[normalem]{ulem}
\usepackage{xcolor, amsmath}
\newcommand\Rey{\mbox{\textit{Re}}}  % Reynolds number
\newcommand\Real{\mbox{Re}}          % cf plain TeX's \Re and Reynolds number

\usepackage{scalerel,stackengine}
\stackMath
\newcommand\reallywidehat[1]{%
\savestack{\tmpbox}{\stretchto{%
  \scaleto{%
    \scalerel*[\widthof{\ensuremath{#1}}]{\kern.1pt\mathchar"0362\kern.1pt}%
    {\rule{0ex}{\textheight}}%WIDTH-LIMITED CIRCUMFLEX
  }{\textheight}% 
}{2.4ex}}%
\stackon[-6.9pt]{#1}{\tmpbox}%
}

\newcommand*\patchAmsMathEnvironmentForLineno[1]{
  \expandafter\let\csname old#1\expandafter\endcsname\csname #1\endcsname
  \expandafter\let\csname oldend#1\expandafter\endcsname\csname end#1\endcsname
  \renewenvironment{#1}
  {\linenomath\csname old#1\endcsname}
  {\csname oldend#1\endcsname\endlinenomath}}
  \newcommand*\patchBothAmsMathEnvironmentsForLineno[1]{
  \patchAmsMathEnvironmentForLineno{#1}
  \patchAmsMathEnvironmentForLineno{#1*}}
  \AtBeginDocument{
  \patchBothAmsMathEnvironmentsForLineno{equation}
  \patchBothAmsMathEnvironmentsForLineno{align}
  \patchBothAmsMathEnvironmentsForLineno{flalign}
  \patchBothAmsMathEnvironmentsForLineno{alignat}
  \patchBothAmsMathEnvironmentsForLineno{gather}
  \patchBothAmsMathEnvironmentsForLineno{multline}
}

\newcommand{\mycomment}[1]{}
\draftfalse

\journalname{Journal of Advances in Modeling Earth Systems (JAMES)}

\begin{document}

%% ------------------------------------------------------------------------ %%
%  Title
%
% (A title should be specific, informative, and brief. Use
% abbreviations only if they are defined in the abstract. Titles that
% start with general keywords then specific terms are optimized in
% searches)
%
%% ------------------------------------------------------------------------ %%

% Example: \title{This is a test title}

\title{Subgrid parameterizations of ocean mesoscale eddies based on Germano decomposition}

\authors{Pavel Perezhogin\affil{1,2}, Andrey Glazunov\affil{2}}

\affiliation{1}{Courant Institute of Mathematical Sciences, New York University, New York, NY, USA}
\affiliation{2}{Marchuk Institute of Numerical Mathematics, Russian Academy of Sciences, Moscow, Russia}

%% Corresponding Author:
% Corresponding author mailing address and e-mail address:

% (include name and email addresses of the corresponding author.  More
% than one corresponding author is allowed in this LaTeX file and for
% publication; but only one corresponding author is allowed in our
% editorial system.)

% Example: \correspondingauthor{First and Last Name}{email@address.edu}

\correspondingauthor{P.A. Perezhogin}{pperezhogin@gmail.com}

%% Keypoints, final entry on title page.

%  List up to three key points (at least one is required)
%  Key Points summarize the main points and conclusions of the article
%  Each must be 140 characters or fewer with no special characters or punctuation and must be complete sentences

% Example:
% \begin{keypoints}
% \item	List up to three key points (at least one is required)
% \item	Key Points summarize the main points and conclusions of the article
% \item	Each must be 140 characters or fewer with no special characters or punctuation and must be complete sentences
% \end{keypoints}

\begin{keypoints}
\item We propose a three-component %new 
subgrid model consistent with the physics of 2D fluids using \citeA{germano1986proposal} decomposition
%We apply \citeA{germano1986proposal} decomposition to propose new subgrid models consistent with the physics of 2D fluids
\item The new subgrid model accurately predicts the spectral transfer of energy and enstrophy and improves a posteriori experiments
\item A backscattering component (Reynolds stress) improves coarse-grid ocean models based on QG and primitive equations
\end{keypoints}

%% ------------------------------------------------------------------------ %%
%
%  ABSTRACT and PLAIN LANGUAGE SUMMARY
%
% A good Abstract will begin with a short description of the problem
% being addressed, briefly describe the new data or analyses, then
% briefly states the main conclusion(s) and how they are supported and
% uncertainties.

% The Plain Language Summary should be written for a broad audience,
% including journalists and the science-interested public, that will not have 
% a background in your field.
%
% A Plain Language Summary is required in GRL, JGR: Planets, JGR: Biogeosciences,
% JGR: Oceans, G-Cubed, Reviews of Geophysics, and JAMES.
% see http://sharingscience.agu.org/creating-plain-language-summary/)
%
%% ------------------------------------------------------------------------ %%

%% \begin{abstract} starts the second page

\begin{abstract}
Ocean models at intermediate resolution ($1/4^o$), which partially resolve mesoscale eddies, can be seen as Large eddy simulations (LES) of the primitive equations, in which the effect of unresolved eddies must be parameterized. In this work, we propose new subgrid models that are consistent with the physics of two-dimensional (2D) flows. We analyze subgrid fluxes in barotropic decaying turbulence using \citeA{germano1986proposal} decomposition. We show that Leonard and Cross stresses are responsible for the enstrophy dissipation, while the Reynolds stress is responsible for additional kinetic energy backscatter. 
%We propose a separate model for each term in the Germano decomposition and refer to the result as a mixed model. 
We utilize these findings to propose a new model, consisting of three parts,  that is compared to a baseline dynamic Smagorinsky model (DSM).
% (mixed model) 
%We utilize these findings to propose new dynamic mixed models and compare them to a baseline dynamic Smagorinsky model (DSM). 
The three-component
%The new mixed 
model accurately simulates the spectral transfer of energy and enstrophy and improves the representation of kinetic energy (KE) spectrum, resolved KE and enstrophy decay in a posteriori experiments. The backscattering component of the new %mixed 
model (Reynolds stress) is implemented both in quasi-geostrophic and primitive equation ocean models and improves statistical characteristics, such as the vertical profile of eddy kinetic energy, meridional overturning circulation and cascades of kinetic and potential energy.
\end{abstract}

\section*{Plain Language Summary}
Ocean models at intermediate resolution contain missing physics term that accounts for the contribution of unresolved mesoscale eddies, which needs to be parameterized. Mesoscale eddies obey complex physics which should be accounted for when proposing a parameterization. Here we consider the interscale transfer of kinetic energy and enstrophy in a barotropic fluid and propose new subgrid models which capture this transfer. Our strategy is to split the subgrid contribution into three parts and propose a model for each term separately. This approach results in excellent a priori performance and improves online simulations. We demonstrate that our analysis of subgrid fluxes generalizes well across flow regimes: the new parameterization of energy redistribution improves barotropic, quasi-geostrophic and primitive equation ocean models.

%\linenumbers

%% main text

\section{Introduction}
The horizontal resolution of the ocean component of  climate models has increased recently from non-eddy resolving resolution (around $1^o$) to  eddy-permitting resolution (around $1/4^o$, \citeA{haarsma2016high}). At this resolution, ocean models do resolve the largest mesoscale eddies but still fail to resolve a substantial part of the mesoscale eddy field. Consequently, such resolutions are often referred to as "grey zone"{}  \cite{hewitt2020resolving}. Classical parameterizations of mesoscale eddies are based on the ideas of Reynolds averaging where temporal or ensemble averaging is used to diagnose the effect of eddies on the mean flow \cite{gent1990isopycnal}. Reynolds averaging is suitable for very coarse ocean models; however, in the grey zone, the Large eddy simulation (LES) approach is preferable  \cite{fox2008can, nadiga2008orientation, graham2013framework, bachman2017scale}. In the LES framework, the effect of unresolved eddies is diagnosed with a spatial filter and referred to as a subgrid forcing \cite{zanna2020data}. This forcing needs to be parameterized with a subgrid model. Recently many new parameterizations of mesoscale eddies were built based on the spatial filtering approach \cite{nadiga2008orientation, frederiksen2012stochastic, san2013approximate, mana2014toward, bachman2017scale, pearson2017evaluation, maulik2016dynamic, maulik2017novel, khani2019diagnosing, khanigradient, bolton2019applications, zanna2020data, guillaumin2021stochastic}.  %However, we note that only a few LES subgrid closures have been implemented in realistic ocean models. %\cite{griffies2000biharmonic, pearson2017evaluation, juricke2020ocean}.

The LES approach has a long history of successful applications in three-dimensional (3D) turbulence \cite{sagaut2006large} and comprises a multitude of methods. The most popular subgrid model is the \citeA{smagorinsky1963general} model which relates the subgrid fluxes to the strain rate tensor.
%A relevant approach to subgrid-scale models 
%used in the LES 
%has been proposed by \citeA{germano1991dynamic}.
This model belongs to a class of so-called "functional models" \cite{sagaut2006large}. Functional models are designed to represent the mean effect of the eddies on the resolved flow. An alternative approach to subgrid modeling is "structural modeling"{} \cite{sagaut2006large}.
%There is an alternative approach to propose a subgrid model -- "structural modeling"{} \cite{sagaut2006large}. 
Structural models utilize formal series expansion to approximate the subgrid forcing. Various approximations of subgrid forcing were proposed over the years: from Velocity gradient models (VGM, \citeA{clark1979evaluation}) to Scale-similarity models (SSM, \citeA{bardina1980improved, bardino1983improved}) and Approximate deconvolution models (ADM, \citeA{stolz2001approximate}). 

A linear combination of structural and functional models is referred to as a "mixed model" \cite{meneveau2000scale}. Mixed models combine the best of both approaches: the structural part provides high correlation with the subgrid forcing and the functional part ensures the numerical stability of the simulations. Such mixed models can be naturally studied in the framework of \citeA{germano1986proposal} decomposition, where the subgrid stress is decomposed into Leonard, Cross and Reynolds stresses. 
Separate functional or structural models for each one of these stress terms are then proposed \cite{horiuti1997new}.
%Separate models (functional or structural) for each one of these stress terms are then proposed \cite{horiuti1997new}. 
We also mention another popular subgrid model in 3D LES: the  "dynamic model"{} of \citeA{germano1991dynamic} which allows the estimation of the eddy viscosity coefficient directly from the resolved flow. 

%A popular approach to reduce the need to tune the eddy viscosity coefficient was proposed by \citeA{germano1991dynamic}.

%Optimal parameters of a subgrid parameterization typically depend on the simulated turbulent flow. \citeA{germano1991dynamic} proposed a "dynamic model"{} allowing for the estimation of the eddy viscosity coefficient directly from the resolved flow. 
%Some new research directions include explicit filtering \cite{bose2010grid}, minimum dissipation models \cite{rozema2015minimum} and algebraic models \cite{xie2020artificial}.

%\citeA{bachman2017scale} evaluated a dynamic model in a primitive equation ocean model and found no clear improvement with respect to a model with a prescribed viscosity coefficient.
%The following models may be of particular interest: estimation of the eddy viscosity coefficient from the resolved flow with the dynamic model of \citeA{germano1991dynamic}. And various approximations of subgrid forcing: Velocity gradient model (VGM, \citeA{clark1979evaluation}), Scale-similarity model (SSM, \citeA{bardina1980improved, bardino1983improved}) and Approximate deconvolution model (ADM, \citeA{stolz2001approximate}). The dynamic model was evaluated in the primitive equation ocean model only once \cite{bachman2017scale}, and its performance was found to be similar to a model with a prescribed viscosity coefficient. 
%We suggest that the performance of LES subgrid models in ocean simulations may be limited by various physical and numerical issues which we discuss below.%, and the main purpose of this study is to reveal these issues.

Quantifying the extent to which ocean models can benefit from the methods developed for 3D LES simulations is an open question. 
%Direct application of these methods to oceanic flows may result in unexpected results. 
For example, subgrid parameterizations in 3D turbulence are mainly suited to simulate energy dissipation by the subgrid eddies \cite{meneveau2000scale}. However, in quasi-2D flows, the energy cascade has an inverse direction \cite{ferrari2009ocean}, and thus subgrid forcing energizes the flow on average. This effect is often referred to as a kinetic energy backscatter (KEB), see \citeA{bachman2018relationship, loose2023diagnosing, thuburn2014cascades, jansen2014parameterizing, grooms2015numerical, zanna2017scale, juricke2020ocean, juricke2023scale, jansen2019toward, bachman2019gme}. 

%The inverse direction of the energy cascade results in additional complications when subgrid models developed for 3D LES are applied to quasi-2D fluids. %For example, the eddy viscosity coefficient related to the inverse energy flux must be negative. 
Dynamic models similar to \citeA{germano1991dynamic} have been proposed for quasi-2D flows, see \citeA{bachman2017scale, pawar2020priori, san2014dynamic, maulik2017dynamic, maulik2017stable}. These models simulate only the forward energy transfer, and consequently, their consistency with the physics of quasi-2D flows is limited. On the contrary, various structural models have been shown to simulate the backward transfer of energy, see for example \citeA{chen2003physical, chen2006physical, bouchet2003parameterization, nadiga2008orientation, mana2014toward, maulik2017novel, anstey2017deformation, zanna2020data, khanigradient}. In this paper, we apply the approach of structural modeling to represent the backward energy transfer and propose new dynamic mixed models.

The existing dynamic models in quasi-2D fluids often suffer from a build-up of energy near the grid scale \cite{bachman2017scale, maulik2017dynamic, maulik2017stable, guan2021stable}. This indicates that numerical effects may lead to large errors even in physically meaningful parameterizations \cite{ghosal1996analysis, chow2003further}. In particular, \citeA{thuburn2014cascades} shows that the subgrid kinetic energy transfer diagnosed from the high-resolution data significantly depends on the choice of the numerical scheme. In this paper, 
%instead of proposing a subgrid model which is consistent with a given numerical scheme, 
we reduce discretization errors by leveraging an explicit filtering approach \cite{gullbrand2003effect, carati2001modelling, winckelmans2001explicit, lund2003use, bose2010grid}. The explicit filtering approach treats a filter width and a grid step as independent parameters. The role of discretization errors can be then reduced by enlarging a filter-to-grid width ratio (FGR, \citeA{bose2010grid, sarwar2017linking}).

The goal of our study is to propose new subgrid momentum closures of ocean mesoscale eddies which are consistent with the physics of quasi-2D flow. We analyze the enstrophy and energy fluxes in barotropic decaying turbulence using \citeA{germano1986proposal} decomposition. We show that the Leonard and Cross stresses describe the enstrophy dissipation, and Reynolds stress describes additional energy backscatter. Leonard stress can be computed directly. We propose a biharmonic Smagorinsky model for the Cross stress and a structural model for the Reynolds stress which is similar to \citeA{horiuti1997new}. We estimate the Smagorinsky coefficient using the dynamic model of \citeA{germano1991dynamic}. The  energy flux produced by backscatter parameterization is determined by considering the budget of subgrid KE \cite{jansen2014parameterizing} and estimation of subgrid KE \cite{khanigradient}. The resulting three-component subgrid model simulates energy and enstrophy fluxes and improves a posteriori experiments. Additionally, we show that the new backscatter model (Reynolds stress) improves quasi-geostrophic and primitive equation ocean models.

The study is structured as follows. In Section \ref{sec:gov_eq} we describe the governing equations. In Section \ref{a_priori_section} we analyze subgrid fluxes using \citeA{germano1986proposal} decomposition. In Section \ref{LES_models} we describe subgrid models. In Section \ref{sec:posteriori} subgrid models are evaluated in a posteriori experiments. Section \ref{sec:implementation} is devoted to the implementation to more realistic ocean models.

%In Section \ref{sec:gov_eq} we  describe an idealized ocean model: decaying barotropic turbulence. In Section \ref{a_priori_section} we analyze subgrid fluxes using \citeA{germano1986proposal} decomposition, and reveal parts of the subgrid stress responsible for the enstrophy dissipation and kinetic energy backscatter. In Section \ref{LES_models} we introduce a baseline dynamic Smagorinsky model (DSM, \cite{germano1991dynamic}), and propose new mixed models which improve the reproducing of the energy and enstrophy transfer spectra. In Section \ref{sec:posteriori} we evaluate the proposed subgrid models a posteriori. In Section \ref{sec:implementation} we implement a new  backscattering model (Reynolds stress) to quasi-geostrophic and primitive equation ocean models and compare it to the \citeA{jansen2014parameterizing} backscatter parameterization.

\section{Governing equations} \label{sec:gov_eq}
In this section, we describe a Direct numerical simulation (DNS) of decaying barotropic turbulence and numerical schemes.

\begin{figure}[h!]
\centerline{\includegraphics[width=1.0\textwidth]{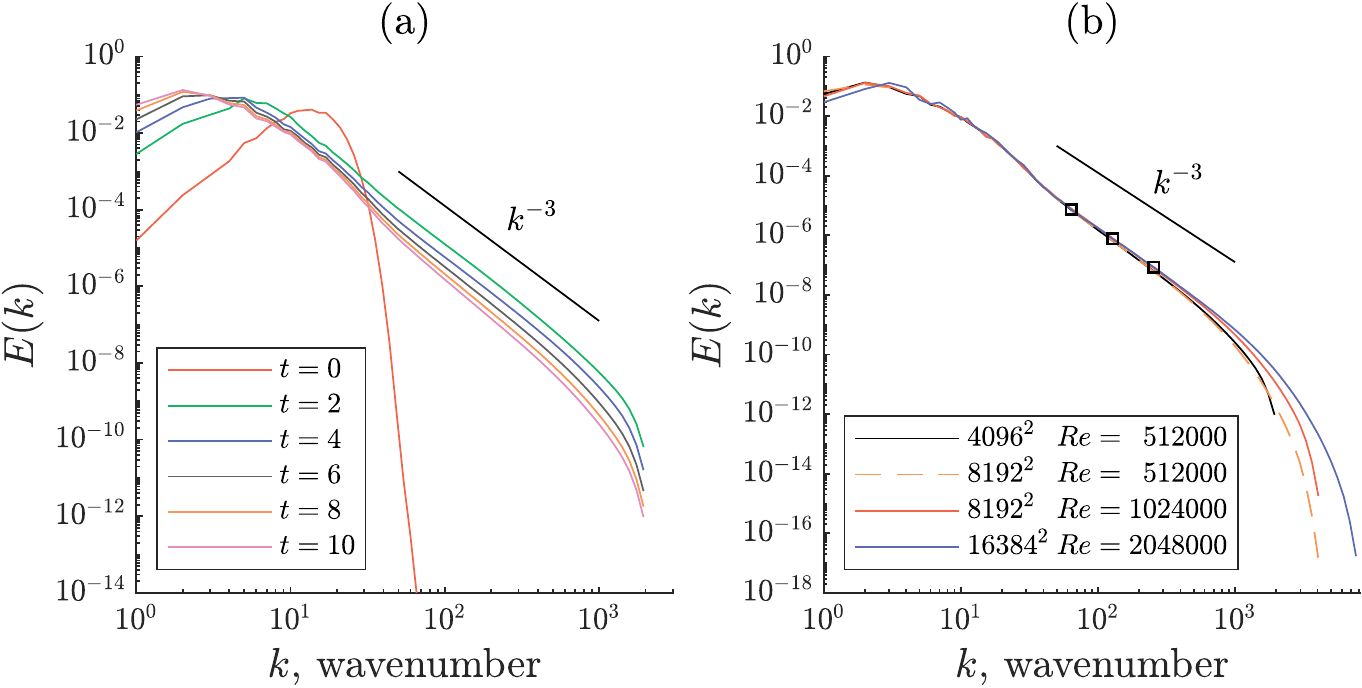}}
\caption{Kinetic energy spectrum in DNS simulations: (a) time evolution at mesh $4096^2$ and $\Rey=512000$, and (b) additional combinations of mesh and Reynolds number at $t=10$. Squares show the cutoff wavenumber  ($\pi/\Delta_g$, where $\Delta_g$ -- grid step) for the coarse LES models at resolutions $128^2$, $256^2$ and $512^2$.}
\label{fig_spectrum_DNS}
\end{figure} 

The dimensionless barotropic vorticity equation in a doubly periodic domain of size $2\pi \times 2\pi$ is \cite{maulik2017dynamic, maulik2017stable, guan2021stable}:
\begin{gather}
	\frac{\partial \omega}{\partial t} + \frac{\partial}{\partial x_j} \left( u_j \omega \right) = \frac{1}{\Rey} \nabla^2 \omega, ~ \nabla^2 \psi = \omega \label{gov_eq},
\end{gather}
where $x_1$ and $x_2$ are Cartesian coordinates, $\nabla = \left(\partial_{x_1}, \partial_{x_2} \right)$ is the gradient operator. We assume summation over the repeated indices ($j=1,2$). The relative vorticity $\omega$, streamfunction $\psi$ and velocity vector components $u_j$ are related to each other as $\omega = \partial_{x_1} u_2 - \partial_{x_2} u_1$ and $\left(u_1, u_2 \right) = \left( - \partial_{x_2} \psi, \partial_{x_1} \psi \right)$. The Reynolds number is defined by dimensional RMS velocity ($\widetilde{u}_{\mathrm{rms}}$), domain size $2\pi \widetilde{L}$ and molecular viscosity ($\widetilde{\nu}$) as $\Rey = \widetilde{u}_{\mathrm{rms}} \widetilde{L} / \widetilde{\nu}$.

The turbulence is initialized with a random divergence-free flow having the following kinetic energy density (per unit wavenumber $k$ and unit area):
\begin{equation}
	E(k) = A k^4 \exp \left(-(k/k_p)^2 \right),~ A = \frac{4 k_p^{-5}}{3 \sqrt{\pi}}, \label{perturbation}
\end{equation}
where $k_p = 10$,  $k = \sqrt{k_1^2+k_2^2}$ and $k_1$,$k_2$ are components of wavevector.
The normalization constant $A$ is chosen to set the RMS velocity to one: $u_{\mathrm{rms}} = \left(2 \int E(k) dk \right)^{1/2} = 1$. We integrate equations \eqref{gov_eq} with initial perturbation of form \eqref{perturbation} until the dimensionless time $t=10$. 

In Figure \ref{fig_spectrum_DNS}(a) we show decay of the kinetic energy spectrum in the DNS simulation for a combination of parameters that we use throughout the paper: resolution $4096^2$ and $\Rey=512000$. The spectrum is averaged over 50 realizations of the initial random field. The chosen Reynolds number is very large, and further increase of $\Rey$ or resolution does not influence significantly the band of scales resolved by the coarse LES models, see squares in Figure \ref{fig_spectrum_DNS}(b).

%All statistical characteristics of the turbulence presented are averaged over an ensemble of 50 realizations of the initial random field.
%Computations are carried out in hydrodynamic code of \citeA{mortikov2019numerical}, particularly in its modification for 2D hydrodynamics \cite{perezhogin2017comparison}. 
%The numerical mesh is uniform.
%The Reynolds number ($\Rey = 512000$) is 4 times larger, and the resolution of DNS ($4096 \times 4096$) is twice that in work \citeA{maulik2017dynamic}.

Both DNS and LES models are discretized with the same second-order numerical scheme, which is a typical choice in realistic ocean models \cite{madec2017nemo, adcroft2019gfdl}. Specifically, we use the Arakawa scheme on the C grid conserving energy and enstrophy
\cite{arakawa1997computational,maulik2017stable} and second-order approximation of the Poisson equation in \eqref{gov_eq} which is solved in Fourier space. A three-stage Runge-Kutta (RK3) scheme \cite{skamarock2008description} is used for time integration, with the time step $\Delta t$ satisfying the  linear stability criterion $\mathrm{CFL}=$ $\Delta t \max_j(|u_j|)/\Delta_g$ $< 0.7$, where $\Delta_g$ is the grid step. 

\section{A priori analysis of the interaction with subgrid eddies} \label{a_priori_section}
In this section, we diagnose the forcing produced by the subgrid eddies on the resolved flow. The analysis of subgrid forcing will guide the development of new subgrid models capable to simulate energy and enstrophy fluxes. We perform the analysis of the subgrid energy budget to propose a parameterization that is energetically consistent, see \citeA{jansen2014parameterizing}. Additionally, we use \citeA{germano1986proposal} decomposition to identify the components of subgrid forcing responsible for the energy and enstrophy fluxes.

\subsection{Filtered equations}
Following the LES approach \cite{sagaut2006large}, we introduce a spatial filter $\overline{(\cdot)}$ decomposing the flow into the resolved part and unresolved or subgrid eddies.  The filter is Gaussian and defined in Fourier space by the transfer function $\exp\left( - \overline{\Delta}^2 k^2 /24 \right)$, where $\overline{\Delta}$ -- filter width. By applying the filter to the governing equations \eqref{gov_eq}, we obtain an equation for the large-scale flow:
\begin{equation}
	\frac{\partial \overline{\omega}}{\partial t} + \frac{\partial}{\partial x_j} \left( \overline{u}_j \overline{\omega} \right) = \frac{1}{\Rey} \nabla^2 \overline{\omega}  - \frac{\partial}{\partial x_j} \left( \sigma_j \right), ~ \nabla^2 \overline{\psi} = \overline{\omega}, \label{LES_eq}
\end{equation}
which is unclosed and contains interaction with the subgrid eddies (subgrid flux):
\begin{equation}
	\sigma_j = \overline{u_j \omega} - \overline{u}_j \overline{\omega}. \label{SFS_stress}
\end{equation}

The spatial filter mimics the effect of a finite resolution and its width should be proportional to the grid step of the coarse LES model ($\Delta_g$). The Gaussian filters related to the coarse resolutions of $128^2$, $256^2$ and $512^2$ points are denoted as $\Delta_{128}$, $\Delta_{256}$ and $\Delta_{512}$, respectively. We set the filter-to-grid width ratio as  $FGR=\overline{\Delta}/\Delta_g=\sqrt{6}$, and explain our choice in Section \ref{sec:posteriori}. Note that a priori analysis is performed on a DNS grid, and the coarse model's grid step is used only to guide the choice of the filter width $\overline{\Delta}$. 

\subsection{Domain-averaged energy exchange with subgrid eddies}

\begin{figure}
	\centerline{\includegraphics[width=\textwidth]{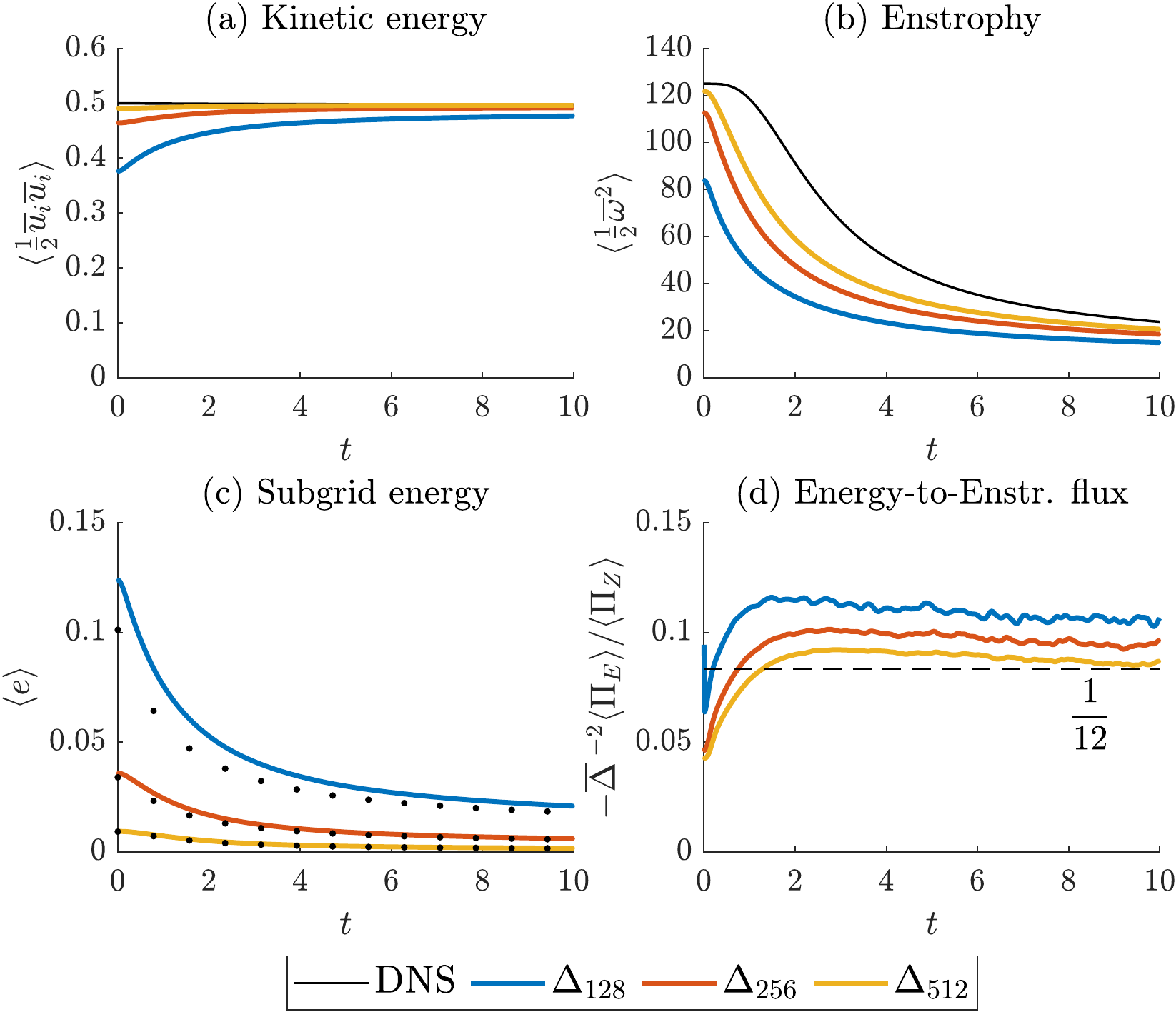}}
	\caption{
		(a) Kinetic energy and (b) enstrophy
		in DNS (black line) and filtered solutions (in colors),
		(c) subgrid energy (solid lines) and its estimation
		according to Eq. \eqref{eq:khani_model}
		in dots,
		(d) the ratio of energy and enstrophy fluxes; the filter is Gaussian with different widths:  $\Delta_{128}>\Delta_{256}>\Delta_{512}$.} \label{a_priori_series}
\end{figure}  
The coarse LES model should simulate the statistical properties of the filtered DNS \cite{frezat2022posteriori}, such as kinetic energy and enstrophy of the filtered solution. The enstrophy of the filtered solution decays (Figure \ref{a_priori_series}(b)), and it is a consequence of the direct enstrophy cascade. On the contrary, the resolved energy in decaying 2D turbulence grows for a very high Reynolds number (Figure \ref{a_priori_series}(a)), and it is a consequence of  the redistribution of the kinetic energy towards large scales (inverse cascade, \cite{kraichnan1967inertial, leith1968diffusion, batchelor1969computation}), see also Figure 9 in \citeA{thuburn2014cascades}.

To accurately simulate the energy of the filtered solution, the subgrid parameterization should predict the energy exchange between resolved flow and subgrid eddies. Consider the budget of subgrid kinetic energy (Eq. (7) in \citeA{jansen2014parameterizing}):
\begin{equation}
    \frac{d}{dt} \langle e \rangle = \langle \Pi_E \rangle - \langle D \rangle, \label{eq:subgrid_KE_equation}
\end{equation}
where $\langle \cdot \rangle$ is the domain-averaging, $e=\frac{1}{2}(\overline{u_i u_i} - \overline{u}_i \overline{u}_i)$ is the subgrid kinetic energy \cite{germano1992turbulence, ghosal1995dynamic}, $D\geq 0$ is the dissipation of subgrid KE, $\Pi_{E} = \sigma_j  \partial_{x_j} \overline{\psi}$ is the energy flux from resolved to subgrid scales. We assume that $D=0$ because there is no bottom drag in governing equation (Eq. \eqref{gov_eq}), see \citeA{jansen2014parameterizing, jansen2019toward}.
%The dissipation of subgrid KE is usually parameterized by frictional dissipation in the bottom boundary layer \cite{jansen2014parameterizing, jansen2019toward}, and thus in our case we assume $D=0$. 
The simplest way to predict the energy flux is to consider a statistically stationary case ($\frac{d}{dt} \langle e \rangle \approx 0$) in equation \eqref{eq:subgrid_KE_equation} which gives zero energy exchange between resolved and subgrid scales  $\langle \Pi_E \rangle \approx 0$, see \citeA{jansen2014parameterizing, thuburn2014cascades}. This approach is not suitable for the simulation of decaying turbulence, which is not stationary. A more accurate approach would include a numerical integration of the equation analogous to \eqref{eq:subgrid_KE_equation} as proposed in \citeA{jansen2015energy}. Our diagnostics show that the subgrid kinetic energy decreases ($\frac{d}{dt} \langle e \rangle < 0$, Figure \ref{a_priori_series}(c)), and thus according to Eq. \eqref{eq:subgrid_KE_equation} it should contribute to the negative subgrid energy flux, i.e. $\langle \Pi_{E} \rangle<0$. That is, subgrid eddies energize the resolved eddies on average. \citeA{partee2022using, khanigradient} proposed a new way to predict the energy of subgrid eddies: it can be estimated given the resolved flow as an alternative to the simulation of Eq. \eqref{eq:subgrid_KE_equation}. The gradient model of \citeA{khanigradient} predicts the subgrid KE using only the resolved flow as:
\begin{equation}
    e = \frac{1}{2} \cdot \frac{\overline{\Delta}^2}{12} \frac{\partial \overline{u}_i}{\partial x_j} \frac{\partial \overline{u}_i}{\partial x_j}, \label{eq:khani_model}
\end{equation}
where we used a standard parameter of the gradient model for the Gaussian filter ($1/12$, \citeA{meneveau2000scale}). In Figure \ref{a_priori_series}(c) we show in black dots that the model (Eq. \eqref{eq:khani_model}) accurately predicts the diagnosed subgrid KE. Using Eq. \eqref{eq:khani_model} and assuming $D=0$, we can estimate the energy flux $\langle \Pi_E \rangle$ from \eqref{eq:subgrid_KE_equation}, where $d/dt$ can be approximated with finite differences. Specifically for 2D decaying turbulence, we can further simplify this method to obtain an interpretable relation between energy and enstrophy fluxes (derived in \ref{appendix:energy_flux}):
\begin{equation}
    \langle \Pi_E \rangle=-\frac{\overline{\Delta}^2}{ 12}\langle \Pi_Z \rangle, \label{eq:Energy_flux_estimate}
\end{equation}
where $\Pi_Z = -\sigma_j \partial_{x_j} \overline{\omega}$ is the enstrophy flux from resolved to subgrid scales. Forward transfer of enstrophy corresponds to a positive flux $\langle \Pi_Z \rangle > 0$. In Figure \ref{a_priori_series}(d) we show that the diagnosed energy and enstrophy fluxes are directed oppositely on average, and the presented estimate of the energy flux (Eq. \eqref{eq:Energy_flux_estimate}) is accurate after the initial adaptation of the turbulence ($t>1$). The formula (Eq. \eqref{eq:Energy_flux_estimate}) will be used to build a new backscatter parameterization.

\subsection{Transfer spectra for Germano decomposition}
\begin{figure}
	\centerline{\includegraphics[width=1.0\textwidth]{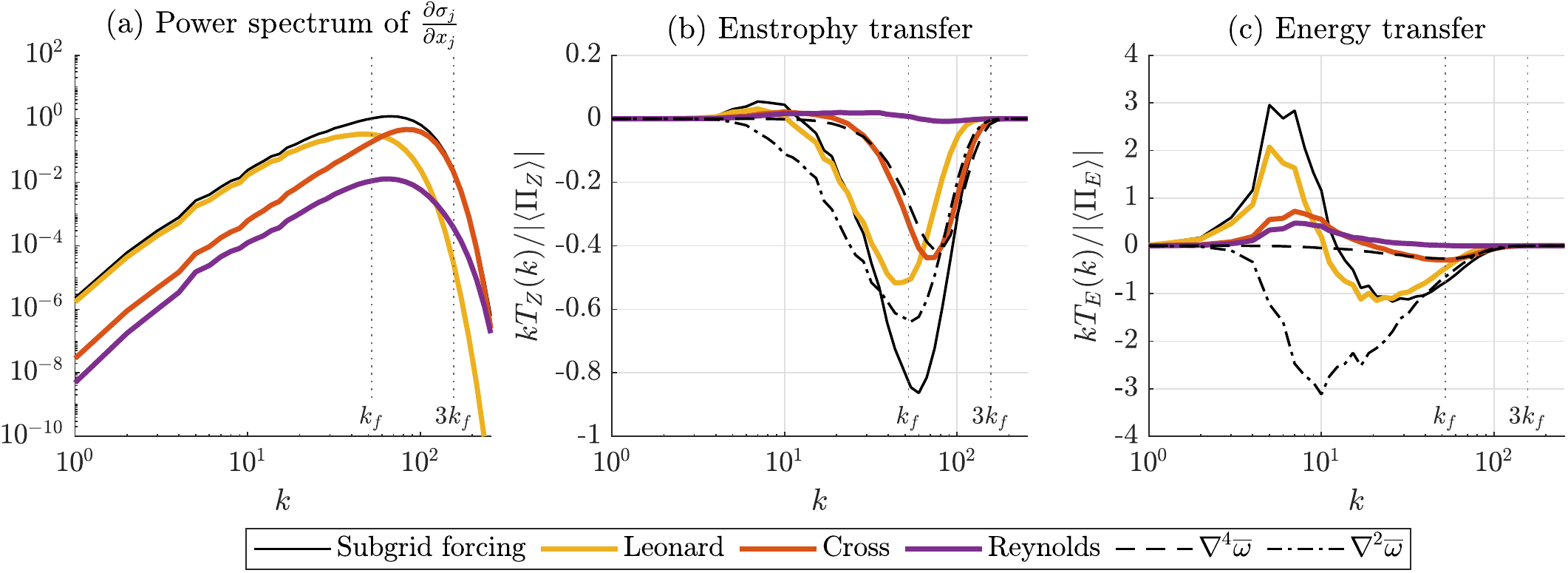}}
	\caption{A priori analysis of the subgrid forcing with Germano decomposition (Eq. \eqref{germano_decomposition}) for the filter with medium width  
 $\Delta_{256}$, $t=2$. (a) Power spectrum of subgrid forcing; (b) enstrophy and (c) energy transfer spectra. The filter scale is defined as $k_f=\pi/\overline{\Delta}$; $\nabla^2\overline{\omega}$ and $\nabla^4\overline{\omega}$ are dissipation spectra produced by laplacian and biharmonic eddy viscosity models.} 
	\label{a_priori_128}
\end{figure}
%In this section, we identify parts of the subgrid forcing responsible for the kinetic energy backscatter and enstrophy dissipation by analyzing spectra of the components of \citeA{germano1986proposal} decomposition.

The subgrid energy and enstrophy transfer spectra are given by, respectively \cite{guan2022learning}:
\begin{gather}
	T_E(k) = \sum_{|\boldsymbol{k}| \in [k,k+1)} \Real\left(  \left(\frac{\partial \sigma_j}{\partial x_j} \right)_{\boldsymbol{k}}^* (\overline{\psi})_{\boldsymbol{k}} \right), \\
 T_Z(k) = \sum_{|\boldsymbol{k}| \in [k,k+1)}
	\Real\left( - \left(\frac{\partial \sigma_j}{\partial x_j} \right)_{\boldsymbol{k}}^* (\overline{\omega})_{\boldsymbol{k}} \right),
\end{gather}
and $(\cdot)_{\mathbf{k}}$ denotes 2D Fourier transform, $(\cdot)^*$ is complex conjugate. These transfer spectra are connected to the energy and enstrophy fluxes ($\Pi_E, \Pi_Z$) as follows:
\begin{equation}
\int T_E(k) \mathrm{d}k = - \langle \Pi_E \rangle, \int T_Z(k) \mathrm{d}k = - \langle \Pi_Z \rangle.
\end{equation}
In Figure \ref{a_priori_128}(b,c) we show the transfer spectra in black line. The subgrid energy and enstrophy transfer contains a small-scale dissipative region ($T_E(k)<0$, $T_Z(k)<0$) and a large-scale backscatter region ($T_E(k)>0$, $T_Z(k)>0$), but the relative contribution of the energy backscatter is higher. We show examples of simple eddy viscosity models (dashed and dot-dashed lines in Figure \ref{a_priori_128}(b,c)). These models are purely dissipative and cannot capture the complex structure of subgrid fluxes. %Thus, we aim to decompose the subgrid forcing into parts to perform a better analysis.

The \citeA{germano1986proposal} decomposition of subgrid vorticity flux is given by \cite{nadiga2008orientation}:
\begin{equation}
	\sigma_j = \underbrace{\overline{\overline{u}_j \overline{\omega}} - \overline{\overline{u}}_j \overline{\overline{\omega}}}_{\text{Leonard}}
	+ \underbrace{\overline{\overline{u}_j \omega'} + \overline{u_j' \overline{\omega}} - \overline{\overline{u}}_j \overline{\omega'}
	-\overline{u'_j}~\overline{\overline{\omega}}}_{\text{Cross}} 
	+ \underbrace{\overline{ u_j' \omega'} - \overline{u'_j}~ \overline{\omega'}}_{\text{Reynolds}}, \label{germano_decomposition}
\end{equation}
where primed quantities denote subgrid eddies, $\omega' = \omega - \overline{\omega}$ and $u_j' = u_j - \overline{u}_j$. The Reynolds stress represents the effect on the resolved flow from eddy-eddy interactions, Cross stress represents the effect of eddy-resolved flow interactions. Finally, the Leonard stress contains only the resolved fields and can be directly computed given $\overline{u}_j$ and $\overline{\omega}$.

In Figure \ref{a_priori_128} we show the spectral content for each component in the Germano decomposition. The enstrophy dissipation is mostly represented by Leonard and Cross stresses, see Figure \ref{a_priori_128}(b). Also, the enstrophy dissipation by the Cross stress can be approximated by the biharmonic viscosity model ($\nabla^4 \overline{\omega}$), see the dashed line in Figure \ref{a_priori_128}(b). These findings will be used to propose a mixed dissipative model of subgrid forcing. The kinetic energy backscatter is influenced by Leonard, Cross and Reynolds stresses, but only the Reynolds stress almost purely represents the positive energy transfer (Figure \ref{a_priori_128}(c)), and this property will be used to propose a new backscatter model. The contribution of Germano decomposition components to the energy and enstrophy transfer is similar for the other filter widths.

\section{Subgrid models}  \label{LES_models}
In this section, we describe the dynamic Smagorinsky model and propose new dissipative and backscattering models by applying the results of a priori analysis.

\subsection{Dynamic Smagorinsky model (DSM)}
The dynamic Smagorinsky model (DSM) is a popular baseline subgrid model in quasi-2D turbulence research \cite{maulik2017novel, pawar2020priori, guan2021stable, frezat2022posteriori}. The Smagorinsky eddy viscosity model is given by:
\begin{equation}
    \sigma_j \approx \sigma_j^{DSM} = - C_S^2 \overline{\Delta}^{\, 2}  |\overline{S}| \frac{\partial \overline{\omega}}{\partial x_j}, \label{eq:Smagorinsky_model}
\end{equation}
where $C_S$ is the Smagorinsky coefficient. Filtered strain-rate tensor is $\overline{S}_{ij}=\frac{1}{2} \left( 
\partial_{x_j} \overline{u}_i + \partial_{x_i} \overline{u}_j \right)$ and its modulus $|\overline{S}|=\sqrt{2 \overline{S}_{ij} \overline{S}_{ij}}$. In the dynamic model of \citeA{germano1991dynamic}, a free parameter ($C_S$) is estimated from the resolved subgrid flux $l_j = \widehat{\overline{u}_j \overline{\omega}} - \widehat{\overline{u}}_j \widehat{\overline{\omega}}$, where a new test filter $\widehat{(\cdot)}$ of width $\widehat{\Delta}$ is introduced. The resolved subgrid flux can be decomposed as follows  (Germano identity):
\begin{equation}
    l_j = \Sigma_j - \widehat{\sigma}_j,  \label{eq:germano_identity}   
\end{equation}
where $\Sigma_j = \widehat{\overline{ u_j \omega}} - \widehat{\overline{u}}_j \widehat{\overline{\omega}}$ is the subgrid flux  with respect to the combined filter $\widehat{\overline{(\cdot)}}$ of width $\widehat{\overline{\Delta}} = \sqrt{\overline{\Delta}^2 + \widehat{\Delta}^2}$ \cite{germano1992turbulence}. Substituting Smagorinsky model \eqref{eq:Smagorinsky_model} to the Germano identity \eqref{eq:germano_identity} and applying least squares procedure of \citeA{ghosal1995dynamic}, we determine the Smagorinsky coefficient:
\begin{equation}
    C_S^2 = \frac{\langle l_j \alpha_j \rangle}{\langle \alpha_j \alpha_j \rangle}, \label{eq:DSM}
\end{equation}
where $\langle  \cdot \rangle$ is the spatial averaging and 
\begin{equation}
    \alpha_j = - \widehat{\overline{\Delta}}^{\, 2}  |\widehat{\overline{S}}| \frac{\partial \widehat{\overline{\omega}}}{\partial x_j} + \reallywidehat{\overline{\Delta}^{\, 2}  |\overline{S}| \dfrac{\partial \overline{\omega}}{\partial x_j}}.
\end{equation}
To reduce the number of free parameters, we set the test filter equal to the base filter, i.e. $\widehat{(\cdot)}=\overline{(\cdot)}$, and consequently $\widehat{\overline{\Delta}}/\overline{\Delta}=\sqrt{2}$. 

\begin{figure}
\centerline{\includegraphics[width=1.0\textwidth]{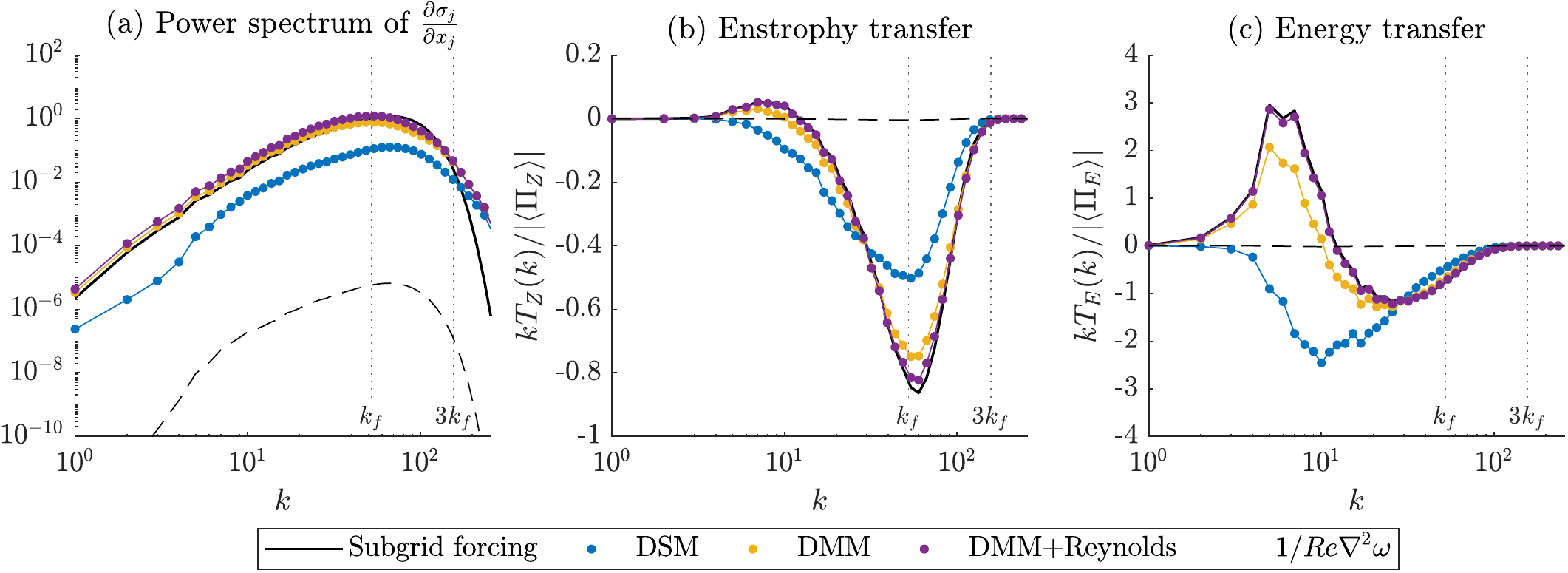}}
	\caption{A priori analysis of subgrid models: DSM is dynamic Smagorinsky model, DMM is dynamic mixed model comprising Leonard stress, DMM+Reynolds includes an additional backscatter parameterization (Reynolds stress). Medium filter width $\Delta_{256}$ and $t=2$. Subgrid models are computed given the filtered DNS data on the grid of DNS. Dashed line shows contribution of  the molecular viscosity at $\Rey=512000$.} \label{fig_SFS_a_priori_spectra}
\end{figure}

The spectral properties of the DSM model (Eq. \eqref{eq:Smagorinsky_model} and \eqref{eq:DSM}) in a priori analysis are shown in Figure \ref{fig_SFS_a_priori_spectra}.  The DSM is a purely dissipative model, and it predicts the enstrophy dissipation of the subgrid forcing reasonably well (Figure \ref{fig_SFS_a_priori_spectra}(b)). However, it introduces the dissipation of kinetic energy on large scales, where the subgrid forcing has a significant positive transfer, i.e. backscatter (Figure \ref{fig_SFS_a_priori_spectra}(c)).
%, but underestimates the power spectrum of subgrid forcing and most importantly introduces dissipation of kinetic energy in a large-scale region where kinetic energy backscatter is expected. 
Thus we conclude that DSM model is inconsistent with the physics of the quasi-2D fluids, and it needs to be modified.

\subsection{Dynamic mixed model (DMM)}
We first leverage the approach of mixed models \cite{meneveau2000scale} to model the dissipation of enstrophy. The classical mixed model combines Leonard stress (also known as the scale-similarity model, SSM, \citeA{bardina1980improved}) with the laplacian Smagorinsky eddy viscosity model \cite{guan2022learning}. However, we have shown in a priori analysis that the enstrophy dissipation is accurately represented by the combination of the Leonard stress with biharmonic eddy viscosity. We utilize this finding in the mixed model as follows:
\begin{equation}
    \sigma_j^{DMM} = \overline{\overline{u}_j \overline{\omega}} - \overline{\overline{u}}_j \overline{\overline{\omega}} + C_S^4 \overline{\Delta}^{\, 4}  |\overline{S}| \frac{\partial (\nabla^2 \overline{\omega})}{\partial x_j}, \label{eq:mixed_model}
\end{equation}
and dynamic procedure to determine the Smagorinsky coefficient:
\begin{equation}
    C_S^4 = \frac{\langle (l_j-h_j) \alpha_j \rangle}{\langle \alpha_j \alpha_j \rangle}, \label{eq:DMM}
\end{equation}
where
\begin{equation}
 \alpha_j =  \widehat{\overline{\Delta}}^{\, 4}  |\widehat{\overline{S}}| \frac{\partial \nabla^2\widehat{\overline{\omega}}}{\partial x_j} - \reallywidehat{\overline{\Delta}^{\, 4}  |\overline{S}| \dfrac{\partial \nabla^2 \overline{\omega}}{\partial x_j}} \text{ and } h_j = \widehat{\overline{\widehat{\overline{u}}_j \widehat{\overline{\omega}}}} - \widehat{\overline{\widehat{\overline{u}}}}_j \widehat{\overline{\widehat{\overline{\omega}}}} - \left(\widehat{\overline{\overline{u}_j \overline{\omega}}} - \widehat{\overline{\overline{u}}_j \overline{\overline{\omega}}} \right),
\end{equation}
see also \citeA{vreman1994formulation} for definition of $h_j$.

The a priori analysis with the DMM model (Eq. \eqref{eq:mixed_model} and \eqref{eq:DMM}) shows an improvement in the enstrophy dissipation spectrum, power spectrum and kinetic energy backscattering in large scales, see Figure \ref{fig_SFS_a_priori_spectra}. However, the positive energy transfer on large scales by the DMM model is clearly underestimated, and it needs to be further modified to account for the missing backscatter.

\subsection{DMM with backscattering part (DMM+Reynolds)}
We have shown in a priori analysis that the Reynolds stress is a promising candidate for an additional backscatter model: it has a small contribution to the enstrophy budget and almost purely represents a positive transfer of kinetic energy. The Reynolds stress cannot be computed given the filtered fields $\overline{\omega}$ and $\overline{u}_j$, but can be approximated as follows:
\begin{equation}
    \overline{ u_j' \omega'} - \overline{u'_j}~ \overline{\omega'} \approx \sigma^{KEB}_j = \overline{\overline{u_j'}~ \overline{\omega'}} - \overline{\overline{u_j'}}~ \overline{\overline{\omega'}}, \label{eq:reynolds_flux}
\end{equation}
where $\overline{u_j'} = \overline{u_j} - \overline{\overline{u_j}}$
and $\overline{\omega'} = \overline{\omega}
 - \overline{\overline{\omega}}$, see \citeA{horiuti1997new} for details. The modification to DMM model accounting for an additional backscatter then reads:
 \begin{equation}
     \sigma_j = \sigma_j^{DMM}+C_R \sigma_j^{KEB},\label{eq:DMM_KEB}
 \end{equation}
 where $\sigma_j^{DMM}$ and its parameter $C_S$ are set in the previous section. The energy balance equation \eqref{eq:Energy_flux_estimate} reads as $\langle \sigma_j \partial_{x_j} \overline{\psi} \rangle =  \frac{\overline{\Delta}^2}{12} \langle \sigma_j \partial_{x_j} \overline{\omega} \rangle$, and allows to choose a free parameter $C_R$ as follows:
 \begin{equation}
     C_R = - \frac{\langle \sigma_j^{DMM} \beta_j \rangle}{\langle \sigma_j^{KEB} \beta_j \rangle}, \label{eq:C_R}
 \end{equation}
where
\begin{equation}
    \beta_j = \frac{\partial \overline{\psi}}{\partial x_j} - \frac{\overline{\Delta}^2}{12} \frac{\partial \overline{\omega}}{\partial x_j}.
\end{equation}

The proposed DMM+Reynolds model (Eq. \eqref{eq:DMM_KEB} and \eqref{eq:C_R}) demonstrates excellent a priori results: it is same good as the DMM model in reproducing the power spectrum and enstrophy dissipation (Figure \ref{fig_SFS_a_priori_spectra}(a,b)), but additionally improves kinetic energy backscatter on large scales (Figure \ref{fig_SFS_a_priori_spectra}(c)). 

\begin{figure}
\centerline{\includegraphics[width=0.5\textwidth]{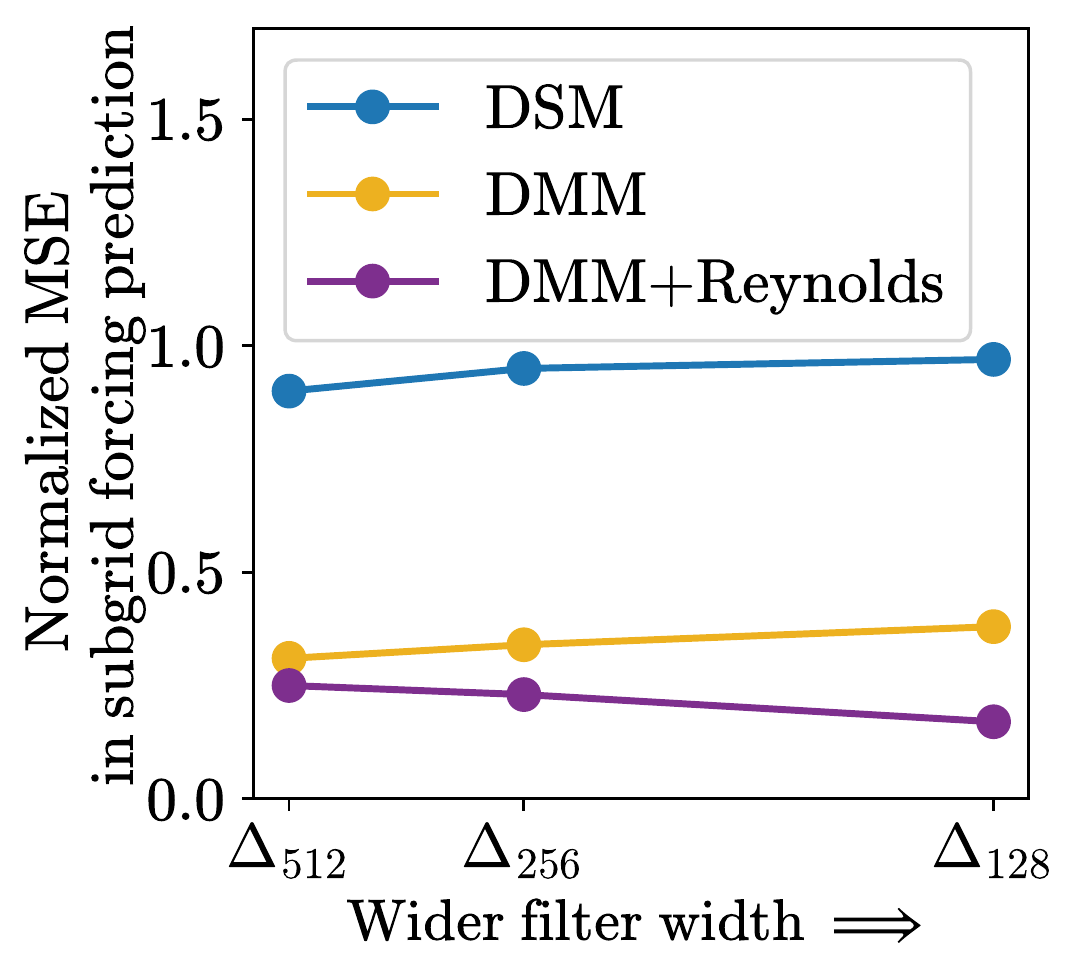}}
	\caption{Mean squared error ($\mathrm{MSE}$) in a priori analysis of subgrid models on DNS grid, averaged over $t\in[2,10]$. Error at a single time is: $\mathrm{MSE} = \left\langle \left( \partial_{x_j} \sigma_j
		- \partial_{x_j} \sigma_j^m \right)^2 \right\rangle / 
		\left\langle \left( \partial_{x_j} \sigma_j \right)^2\right\rangle$, where $\sigma_j$ is the subgrid flux and $\sigma_j^m$ is a subgrid model.} \label{fig:MSE_SFS}
\end{figure}

The proposed modifications to the dynamic Smagorinsky model (DMM and DMM+Reynolds) significantly improve the MSE error in the prediction of subgrid forcing, see Figure \ref{fig:MSE_SFS}. We emphasize that the improvement due to including the parameterization of Reynolds stress increases as the filter gets wider, which is expectable because subgrid and Reynolds stresses should be equal for a very large filter width \cite{sullivan2003structure}.

%The numerical schemes used in the LES models are the same  as in the DNS model and they are described in the section \ref{a_priori_section}.
\subsection{Numerical discretization of subgrid models}
We discretize the subgrid models (DSM, DMM and DMM+Reynolds) with the second-order numerical schemes. The spatial Gaussian filter is implemented in Fourier space if $\epsilon=\overline{\Delta}/\Delta_g > \sqrt{6}$ and using second-order approximation otherwise \cite{sagaut1999discrete}:
\begin{equation}
	\overline{\phi} = \frac{1}{24} \epsilon^2 (\phi_{j+1} + \phi_{j-1})
	+ (1 - \frac{\epsilon^2}{12}) \phi_j, \label{discrete_filter}
\end{equation}
where $j$ is an index of a grid node in one direction. The two-dimensional discrete filter is given by a sequential application of one-dimensional filters \eqref{discrete_filter} along $x_1$ and $x_2$, i.e. filter product, see \citeA{sagaut1999discrete}. A combination of filters $\widehat{\overline{(\cdot)}}$ is given by a sequential application of the base and test filters. The only tunable parameter remains in the coarse LES models: filter-to-grid width ratio $\overline{\Delta}/\Delta_g$, and we discuss it in the next section.

\section{A posteriori experiments} \label{sec:posteriori}
In this section, we implement the proposed subgrid models into the LES equation \eqref{LES_eq}, and perform a posteriori experiments. The goal for LES models is to reproduce filtered DNS (fDNS) data on a coarse grid.

\begin{figure}
	\centerline{\includegraphics[width=1\textwidth]{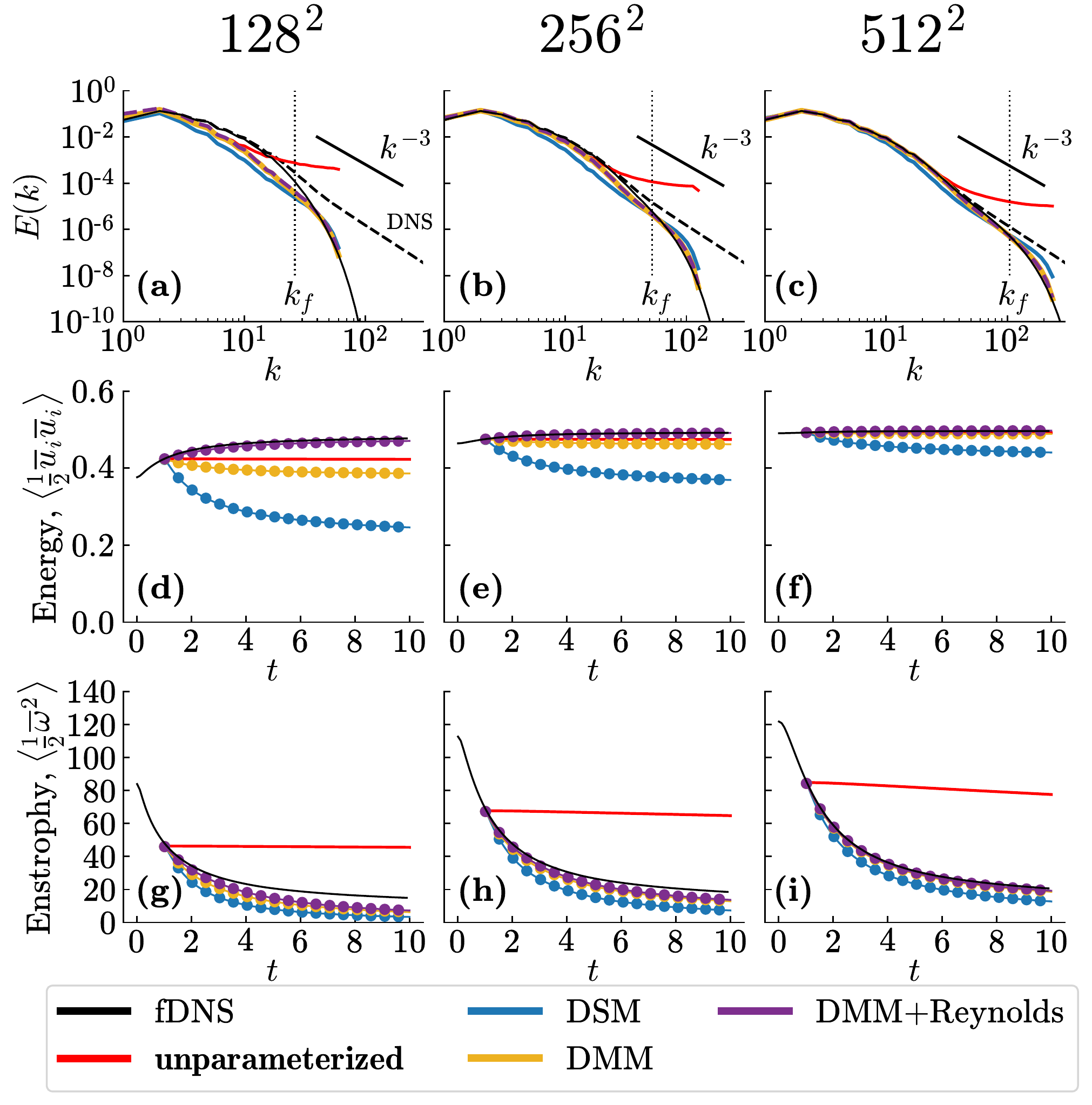}}
	\caption{A posteriori experiments with subgrid models integrated with zero viscosity ($\frac{1}{\Rey}\nabla^2 \overline{\omega}=0$); uparameterized simulation ($\sigma_j=0$) shows the dissipation introduced by the time integration scheme. Upper row: spectrum of KE at $t=10$, middle row: kinetic energy, bottom row: enstrophy. DNS at resolution $4096^2$ and $\Rey=512000$ is used as a reference solution. 
 }
\label{fig:spectrum_energy_enstrophy}
\end{figure}

\subsection{Comparison of subgrid models}
As a reference solution, we use DNS at resolution $4096^2$ and $\Rey=512000$. In order to demonstrate that the proposed subgrid models do not generate numerical noise, we integrate LES equation \eqref{LES_eq} on a coarse grid without molecular viscosity  ($\frac{1}{\Rey}\nabla^2 \overline{\omega}=0$). Note that results with molecular viscosity are almost identical. We also provide simulations with unparameterized model ($\sigma_j=0$), where the only dissipation is related to the time integration scheme (RK3). Neglecting molecular viscosity is a common practice in realistic ocean models, and in our case it is justified by its low impact on scales of the coarse LES models, see Figures \ref{fig_spectrum_DNS} and \ref{fig_SFS_a_priori_spectra}.

Every experiment is computed for an ensemble of 50 realizations. Numerical integration starts at $t=1$, and the initial condition is prepared from DNS data as follows. We first apply a Gaussian filter of width $\overline{\Delta}$ to DNS fields, and then perform spectral truncation of wavenumbers  $|k_i| > \pi/\Delta_g$, where $\Delta_g$  is the grid step of a coarse LES model. We run a posteriori experiments for three resolutions ($128^2$, $256^2$, $512^2$) at a fixed FGR: $\overline{\Delta}/\Delta_g=\sqrt{6}$. This parameter was chosen based on the sensitivity studies and corresponds to a tradeoff between the strength of discretization errors and the number of directly resolved turbulent eddies, see \ref{sec:explicit_filtering}.

All the proposed dynamic models (DSM, DMM, DMM+Reynolds) produce numerically stable solutions without build-up of energy spectrum near the grid scale for a range of resolutions, see upper row in Figure \ref{fig:spectrum_energy_enstrophy}. The mixed models (DMM and DMM+Reynolds) are superior to the baseline DSM. They better reproduce the shape of the energy spectrum of fDNS near the filter scale ($k_f$) and in middle scales, see Figure \ref{fig:spectrum_energy_enstrophy}(a,b,c). The DSM model dissipates too much energy, and DMM model allows to reduce the dissipation. The inclusion of the Reynolds stress introduces the kinetic energy backscatter which allows to simulate the growth of the kinetic energy (Figure \ref{fig:spectrum_energy_enstrophy}(d,e,f)). Note that the effect of the Reynolds stress is enlarging for coarser resolutions consistently with a priori analysis. The decay of enstrophy is improved with the use of new mixed models compared to the baseline DSM (Figure \ref{fig:spectrum_energy_enstrophy}(g,h,i)).%, although we observe too large dissipation at the coarsest resolution, which we discuss in the next section.

\begin{figure}
\centerline{\includegraphics[width=1.0\textwidth]{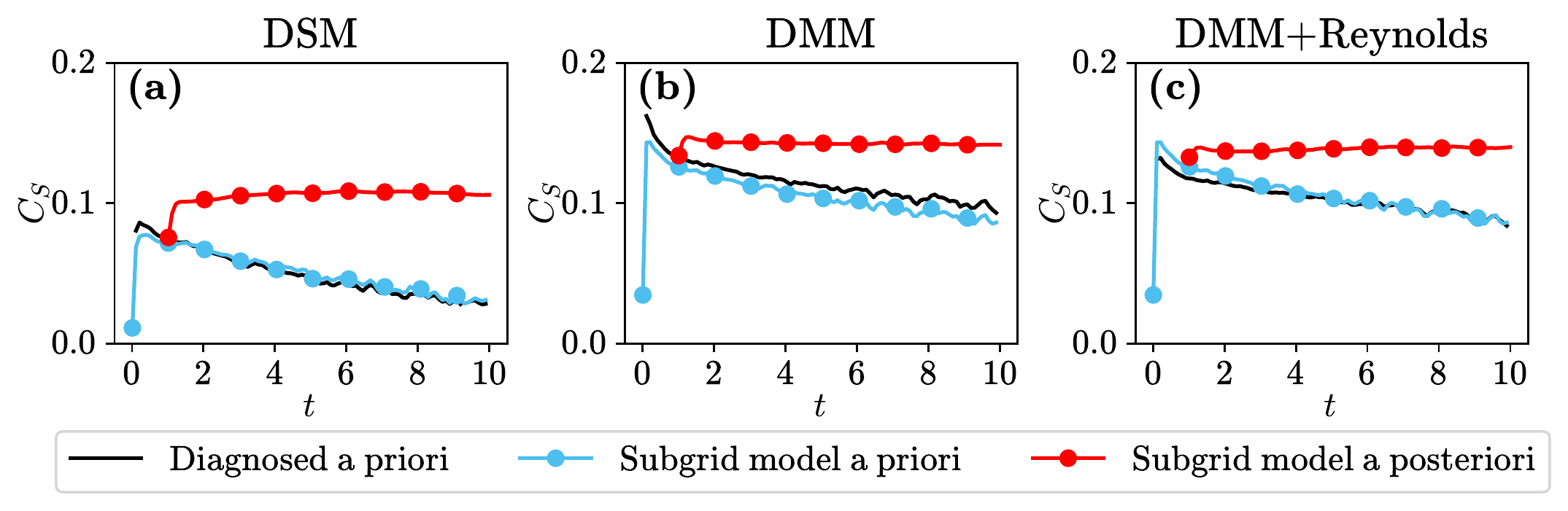}}
	\caption{Comparison of the predicted and diagnosed Smagorinsky coefficient ($C_S$). Black line: $C_S$ diagnosed from DNS data, blue line: prediction of $C_S$ by dynamic model on filtered DNS data, red line: prediction by dynamic model in a posteriori experiment. All experiments have $\Rey=512000$ and filter width $\Delta_{128}$.
	} \label{fig_SFS_a_priori_cscr}
\end{figure} 

\subsection{Scale invariance}
The scale similarity of the kinetic energy spectrum in DNS breaks at the largest scales when ever-enlarging eddies condensate and result in a spectrum different from the power law. Once the eddy scale falls within the subgrid range, all subgrid models fail to reproduce the complicated shape of fDNS spectrum and predict a power law, see Figure \ref{fig:spectrum_energy_enstrophy}(a). For example, the subgrid model DMM+Reynolds energizes the flow, but additional energy resides in the largest scales almost without improving the middle ones compared to DMM. An excessive dissipation of enstrophy (Figure \ref{fig:spectrum_energy_enstrophy}(g)) also indicates that the middle scales are too damped. In this section, we investigate the influence of the scale invariance on the performance of subgrid models.

Dynamic subgrid models are built on the assumption that the Smagorinsky coefficient is scale-invariant, i.e. it is independent of the filter width. However, this assumption violates whenever we deal with a break of self-similarity of the energy spectrum. For this case, a scale-dependent dynamic model was proposed \cite{meneveau1997dynamic, porte2000scale}. Likewise, the scale invariance of the Smagorinsky model can be violated for quasi-2D flows exhibiting the enstrophy cascade, and for this case \citeA{leith1996stochastic} proposed a new eddy viscosity model \cite{fox2008can,bachman2017scale}. A break of the scale invariance of the Smagorinsky model can potentially lead to an inaccurate prediction of the Smagorinsky coefficient by the dynamic procedure of \citeA{germano1991dynamic}. 

In Figure \ref{fig_SFS_a_priori_cscr}(a), we show in black line the Smagorinsky coefficient ($C_S$) diagnosed from the DNS data by the least squares fit of the subgrid flux $\sigma_j$:
\begin{equation}
    C_S^2 = \frac{\langle \sigma_j \alpha_j \rangle}{\langle \alpha_j \alpha_j \rangle}, ~ \alpha_j = - \overline{\Delta}^2 |\overline{S}| \frac{\partial \overline{\omega}}{\partial x_j}.
\end{equation}
The subgrid model DSM applied to filtered DNS data (Eq. \eqref{eq:DSM}) accurately predicts the diagnosed parameter $C_S$, see blue line in Figure \ref{fig_SFS_a_priori_cscr}(a). However, once we evaluate the subgrid model a posteriori, the Smagorinsky coefficient abruptly increases (red line in Figure \ref{fig_SFS_a_priori_cscr}(a)) and it results in the excessive dissipation of enstrophy. We conclude that the scale invariance of the eddy viscosity model has a minor effect on the accuracy of the dynamic procedure, but the main difficulty is in the lack of consistency between a priori and a posteriori performance of the same subgrid model \cite{ross2023benchmarking}. 

In Figure \ref{fig_SFS_a_priori_cscr}(b,c) we show similar difference between a priori and a posteriori results for DMM and DMM+Reynolds models. Note that diagnosed $C_S$ is slightly different for these models, because the least squares fit for DMM+Reynolds model includes additional parameter $C_R$. The diagnosed and predicted values of this parameter are close to $C_R\approx20$, and we do not show it.

\begin{figure}
\centerline{\includegraphics[width=1.0\textwidth]{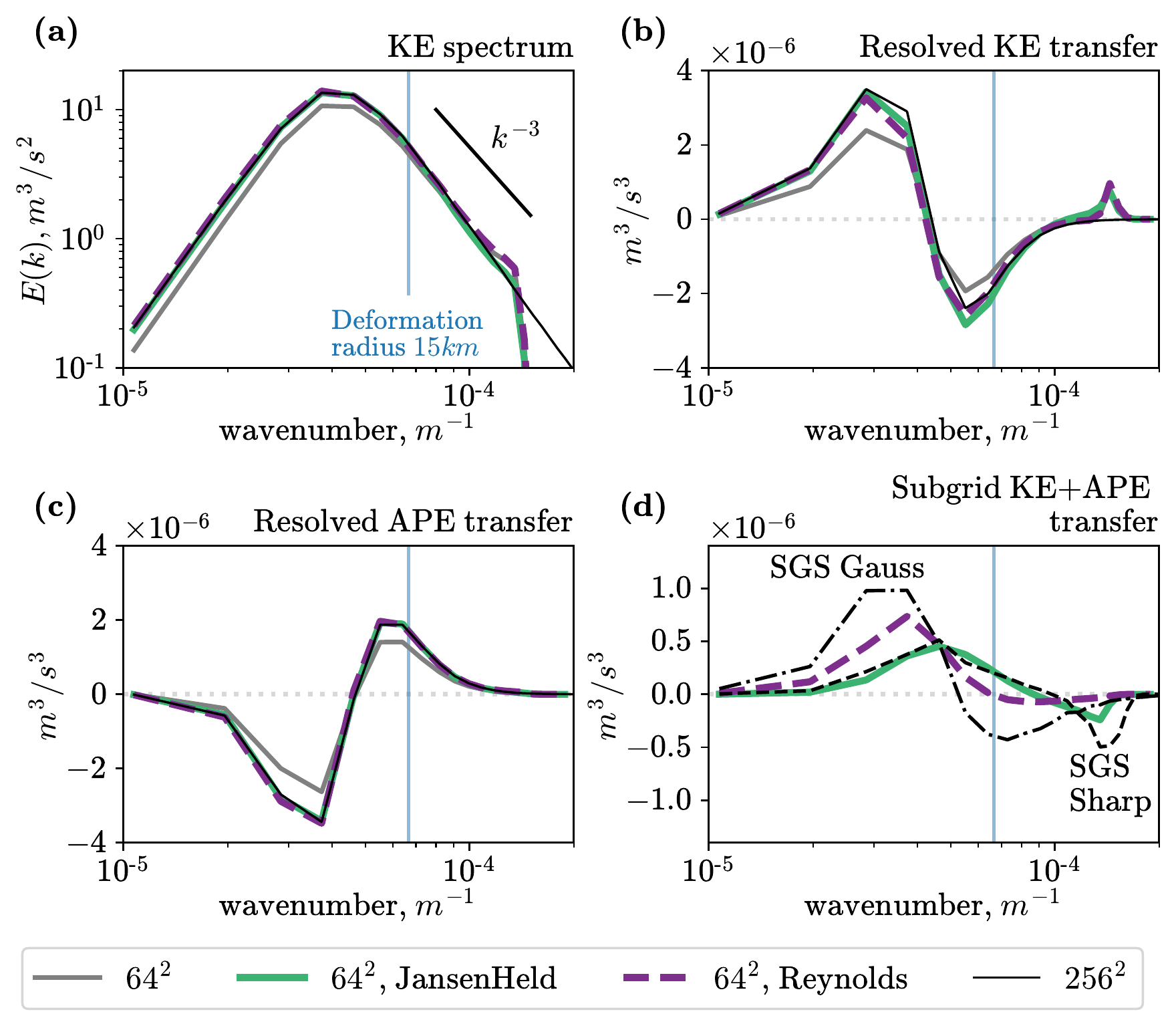}}
	\caption{Experiments in two-layer QG idealized ocean model. High-resolution simulation ($256^2$) is compared to models at coarse resolution, where $64^2$ denotes unparameterized simulation, JansenHeld denotes negative viscosity backscatter, and Reynolds is given in Eq. \eqref{eq:reynolds_flux_pyqg}. (a) Kinetic energy spectrum, (b) resolved transfer of KE, (c) resolved transfer of APE, (d) transfer of total energy by subgrid parameterizations. }%Dashed and dot-dashed lines on panel (d) show the subgrid transfer diagnosed a priori.}
    \label{fig:pyqg}
\end{figure} 

\section{Implementation to QG and primitive equation ocean models} \label{sec:implementation}
We suggest that a similar analysis should be performed in a more realistic set of equations before dynamic mixed parameterizations can be successfully used in ocean simulations. An important outcome of our study is to show that the Reynolds stress can be used as a backscatter parameterization. This is demonstrated in this section in experiments in quasi-geostrophic (QG) and primitive equation ocean models.
% The most simple and new outcome of our study is that the Reynolds stress can be used as a backscatter parameterization, and in this section, we confirm our finding in a posteriori experiments in quasi-geostrophic (QG) and primitive equation ocean models.

\subsection{Two-layer QG model}
We use an idealized QG ocean model (pyqg, \citeA{pyqg}). Our configuration is called "eddy"{} and described in \citeA{ross2023benchmarking,perezhogin2023generative}. The model has two fluid layers in a doubly-periodic domain. It is forced by the prescribed vertical shear of a zonal flow and loses its energy by frictional dissipation in the bottom layer. The numerical scheme is pseudospectral with a highly scale-selective dissipation, which removes enstrophy and numerical noise near the grid scale. 

We extend the Reynolds model (Eq. \eqref{eq:reynolds_flux}) to simulate the subgrid flux of potential vorticity (PV) as follows:
\begin{equation}
    \frac{\partial q}{\partial t} = \cdots - C_R\frac{\partial}{\partial x_j} \left(
    \overline{u_j'~ q'} - \overline{u_j'}~ \overline{q'}
    \right), \label{eq:reynolds_flux_pyqg}
\end{equation}
where $q$ and $u_j$ are the resolved PV and velocity on a coarse grid; $q'=q-\overline{q}$ and $u_j'=u_j-\overline{u_j}$. Note that we omitted one filtering operation in \eqref{eq:reynolds_flux_pyqg} for clarity of numerical implementation (see Section 1.3 in \citeA{layton2012approximate}). The filter $\overline{(\cdot)}$ is Gaussian with $\overline{\Delta}/\Delta_g=2$. The parameterization is applied layerwise with the same $C_R$ which controls the strength of energy injection. We found an optimal value $C_R=7$ by matching the KE spectrum on large scales for high-resolution and coarse parameterized models. 

We choose the parameterization of \citeA{jansen2014parameterizing, jansen2015energy} as a baseline subgrid model. Implementation details are provided in \citeA{ross2023benchmarking}, and we choose the optimal parameters of the parameterization from this paper. JansenHeld subgrid model consists of two parts: small-scale dissipation parameterized by biharmonic viscosity and larger-scale backscatter parameterized by laplacian operator with negative viscosity. 

In Figure \ref{fig:pyqg} we compare coarse models on a grid $64^2$ (grid step $15.6\mathrm{km}$) to the high-resolution simulation ($256^2$, $3.9\mathrm{km}$) after reaching statistical equilibrium: we average the results between $5$ and $20$ years of the simulation for an ensemble of 10 members. The energy cycle comprises two cascades \cite{salmon1978two, vallis2017atmospheric}. The available potential energy (APE) is redistributed towards smaller scales following the direct cascade (Figure \ref{fig:pyqg}(c)), where it is converted to the kinetic energy near the Rossby deformation radius. The kinetic energy is redistributed towards larger scales following the inverse cascade (Figure \ref{fig:pyqg}(b)). The coarse model fails to simulate the energy transfer, and its KE spectral density is smaller compared to the high-resolution model (Figure \ref{fig:pyqg}(a)). Both backscatter parameterizations (Reynolds and JansenHeld) simulate the energy injection in the large scales (Figure \ref{fig:pyqg}(d)) and amplify the resolved (i.e., unparameterized) cascades of KE and APE (Figure \ref{fig:pyqg}(b,c)), which results in a significant improvement in the reproducing of the kinetic energy spectrum. We note that the Reynolds parameterization almost does not change the shape of the KE spectrum near the grid scale and predominantly affects the largest scales. In Figure \ref{fig:pyqg}(d) we compare subgrid models in a posteriori experiments to the subgrid forcing diagnosed a priori, see \citeA{perezhogin2023generative}. The negative viscosity parameterization (JansenHeld) is suitable for the Sharp filter (see  \citeA{kraichnan1976eddy} for explanation), and the Reynolds parameterization is closer to the subgrid forcing diagnosed with the Gaussian filter.

\begin{figure}
\centerline{\includegraphics[width=1.0\textwidth]{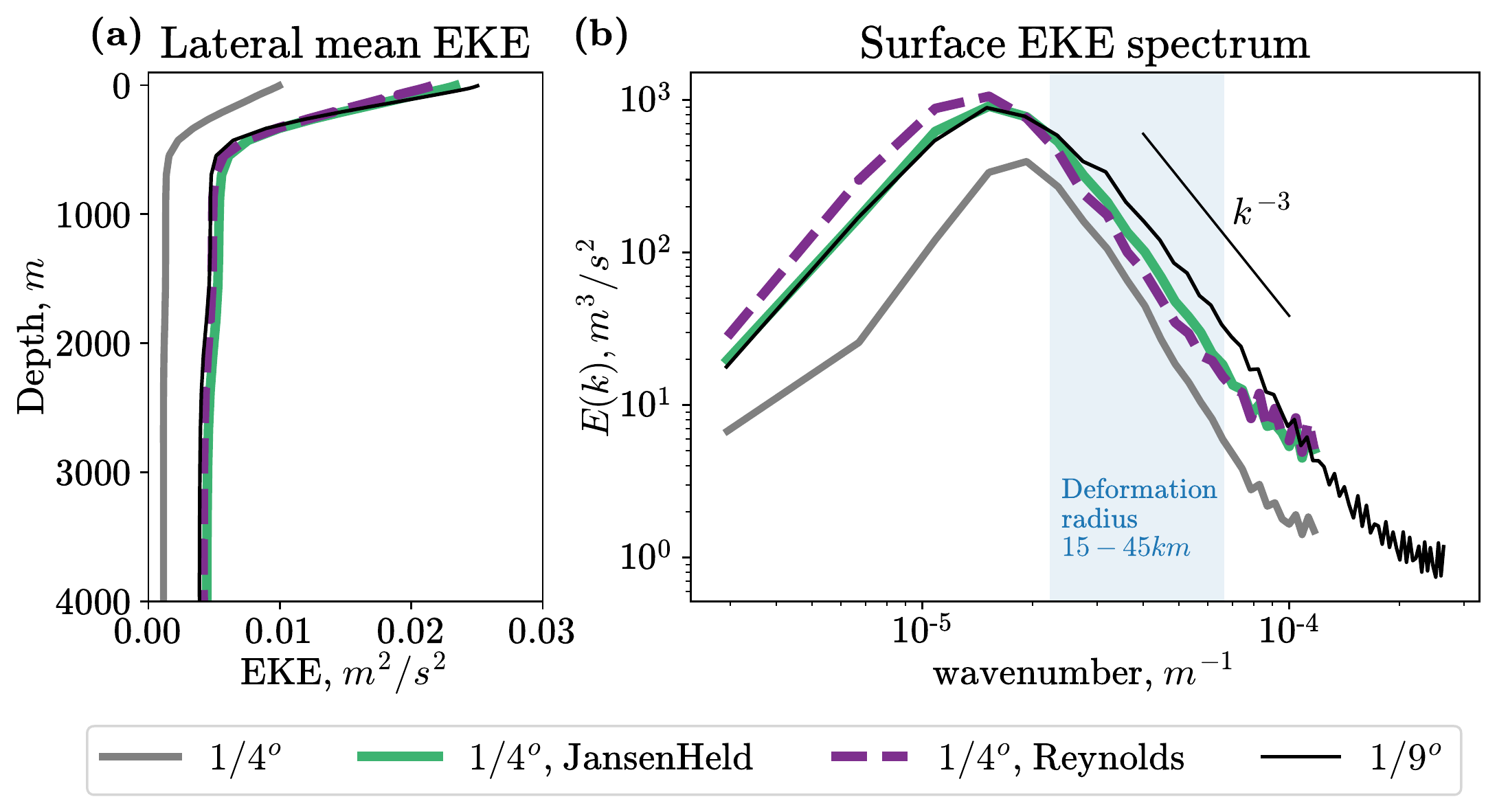}}
	\caption{Experiments in NEMO ocean model in Double Gyre configuration. (a) 20-year mean eddy kinetic energy averaged laterally over the whole domain. (b) spatial spectrum of surface eddy kinetic energy (half the power spectrum of velocity deviations from 1-year mean flow). }
    \label{fig:NEMO_EKE}
\end{figure} 

\subsection{Primitive equation ocean model NEMO}
We use the primitive equation ocean model NEMO \cite{madec2017nemo} in the Double Gyre configuration \cite{levy2010modifications}. The model contains 30 vertical layers in a domain with a flat bottom and vertical walls. The circulation is forced by the prescribed wind stress and buoyancy fluxes on the surface; the equation of state is linear and comprises temperature and salinity. The coarse and reference models have the resolutions of $1/4^o$ (grid step $26.5 \mathrm{km}$) and $1/9^o$ ($11.7 \mathrm{km}$), respectively. The coarse model starts from the snapshot of the high-resolution model, spin-up for 10 years, and integrated for 20 more years to collect statistics. The small-scale dissipation is given by the biharmonic viscosity with a constant coefficient unique for each resolution, see \citeA{perezhogin2020testing} for model parameters. We use a similar baseline subgrid model as in QG simulation: an analog of \citeA{jansen2014parameterizing} backscatter parameterization implemented by the author \cite{perezhogin2019deterministic, perezhogin2020testing}. We use the optimal parameters of the parameterization from these papers.

We extend the Reynolds parameterization (Eq. \eqref{eq:reynolds_flux}) to simulate the subgrid momentum flux as follows:
\begin{equation}
    \frac{\partial u_i}{\partial t} = \cdots - \frac{\partial}{\partial x_j} \Bigg( C_R \left( \overline{u_i' u_j'} - \overline{u_i'} ~\overline{u_j'} \right) \Bigg), ~ i,j \in \{1,2\}, \label{eq:reynolds_primitive}
\end{equation}
where $u_i$ is the resolved horizontal velocity on a coarse grid and $u_i'=u_i - \overline{u_i}$. The parameterization is applied layerwise with the same $C_R$. We observed that if $C_R$ is tuned to improve the kinetic energy, the wider filter $\overline{(\cdot)}$ allows choosing the lower parameter $C_R$. Consequently, we define the filter $\overline{(\cdot)}$ as 2 iterations of the 3-point filter (Eq. \eqref{discrete_filter}) with maximum allowable $\epsilon=\sqrt{6}$. The 3-point filter imposes physical boundary conditions on the velocity: no-normal flow and free slip. To avoid setting boundary conditions on momentum flux, we assume commutation of filter and derivative and compute the first part of the parameterization as follows: $\frac{\partial}{\partial x_j} \Bigg( C_R \overline{u_i' u_j'} \Bigg) \rightarrow \overline{ \frac{\partial}{\partial x_j} \Bigg( C_R u_i' u_j' \Bigg)}$. In preliminary experiments, we realized that it is important to reduce the influence of the boundaries. Therefore, we attenuate the strength of the parameterization smoothly in the vicinity of the wall ($l\leq L$) as follows: $C_R \rightarrow C_R \cdot (1-\cos(\pi l / L)) / 2$, where $l$ is the distance to the wall and $L$ is the length scale of attenuation. After some tuning, we set $L$  as 4 grid steps. The only free parameter $C_R=30$ was chosen from energetic considerations: to obtain the best RMSE in the vertical profile of eddy kinetic energy (EKE).

\begin{table}
	\begin{center}
		\begin{tabular}{l|cccc}
			   RMSE      & SSH, $m$ & SST, $^o C$ & SSS, $psu$ \\ [3pt]
			$1/4^o$     	   & 0.108       & 0.647       & 0.128       \\
			$1/4^o$, JansenHeld     & 0.06 ($-44.16\%$)       & 0.404 ($-37.6\%$)       & 0.112 ($-12.53\%$)      \\
			$1/4^o$, Reynolds & 0.061 ($-43.64\%$)       & 0.321 ($-50.38\%$)       & 0.081 ($-36.38\%$)      
		\end{tabular}
		\caption{The root mean squared errors (RMSE) in 20-year mean sea surface hight (SSH), sea surface temperature (SST) and sea surface salinity (SSS). The error is computed w.r.t. $1/9^o$ model.
		}
		\label{tab:RMSE_NEMO}
	\end{center}
\end{table}

\begin{figure}
\centerline{\includegraphics[width=1.0\textwidth]{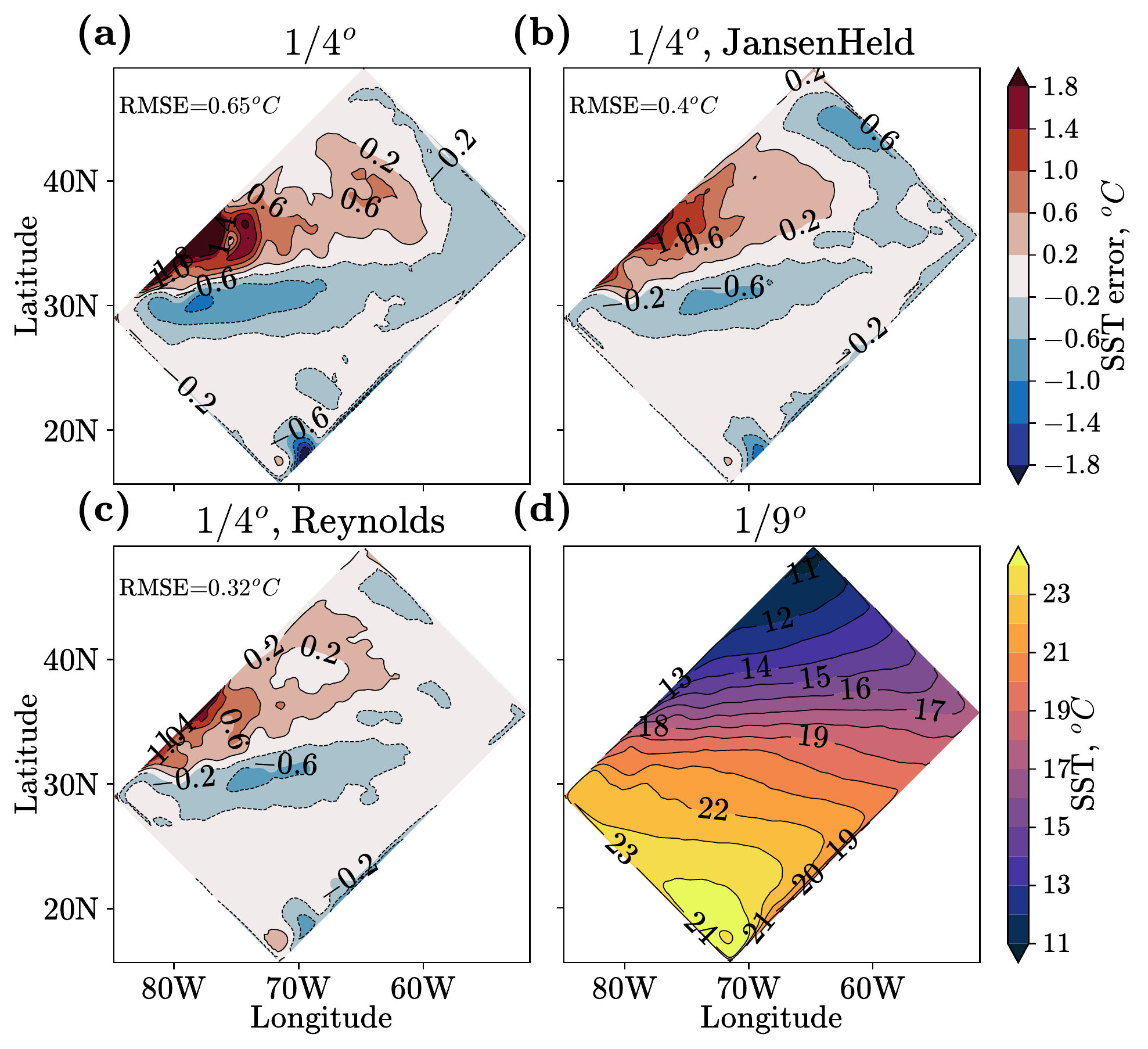}}
	\caption{Experiments in NEMO ocean model. (d) 20-year mean sea surface temperature (SST) in the high-resolution model; (a), (b), (c) errors in SST for coarse models. The Reynolds-parameterized model in panel (c) is given in Eq. \eqref{eq:reynolds_primitive}.}
    \label{fig:NEMO_SST}
\end{figure}

\begin{figure}
\centerline{\includegraphics[width=1.0\textwidth]{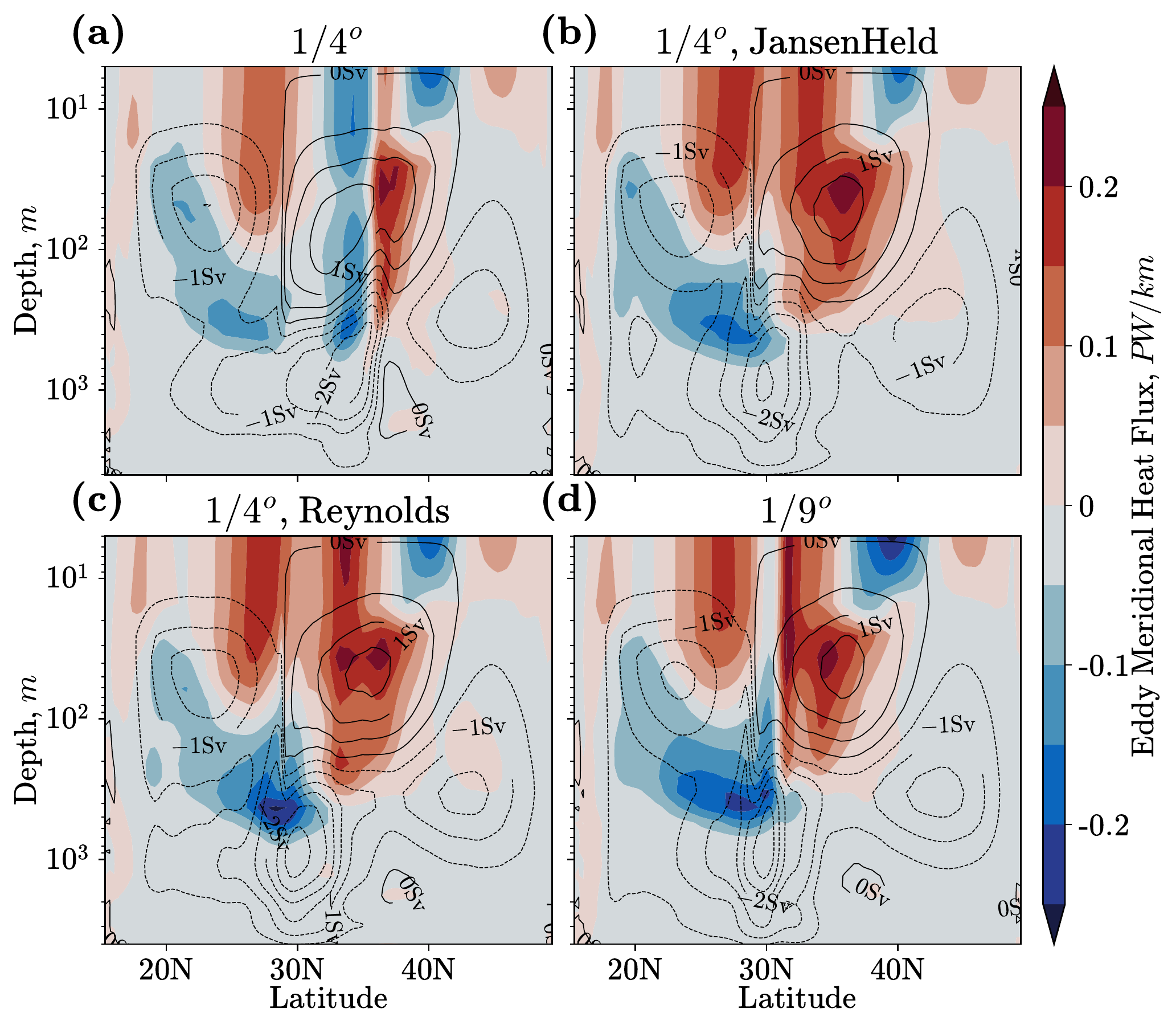}}
	\caption{Experiments in NEMO ocean model. In contours: MOC streamfunction computed as $\Psi_{MOC} = \int_{-H}^z \int_{\mathrm{west}}^{\mathrm{east}} \langle V \rangle_t dxdz$ in Sverdrups, where $\langle V \rangle_t$ -- 20-year mean meridional velocity. In color: the resolved meridional eddy heat flux, defined as zonal integral of $\rho_0 C_p (\langle TV \rangle_t - \langle T \rangle_t \langle V \rangle_t)$, see \citeA{perezhogin2020testing}.}
    \label{fig:NEMO_MOC}
\end{figure}

The Reynolds model works as a backscatter parameterization and energizes the flow on a coarse grid. Similarly to the JansenHeld subgrid model, the EKE can be increased near the surface and the bottom, see Figure \ref{fig:NEMO_EKE}(a). The spectrum of EKE indicates an increase of energy density in all resolved scales for both backscatter parameterizations, see Figure \ref{fig:NEMO_EKE}(b). An improvement in the representation of the resolved eddy activity results in an improvement in several other metrics. In Figure \ref{fig:NEMO_SST} we show 20-year mean sea surface temperature (SST) for the reference simulation and errors for coarse models. The largest error for the unparameterized model is concentrated near the western boundary (Figure \ref{fig:NEMO_SST}(a)) and is explained by the misrepresentation of the western boundary current (WBC, \citeA{levy2010modifications}). Both backscatter parameterizations improve the mean SST near the western boundary (Figure \ref{fig:NEMO_SST}(b,c)), but the Reynolds model is also better in the northern region (Figure \ref{fig:NEMO_SST}(c)). The RMSE in surface fields for temperature and salinity indicates lower errors for the Reynolds model, see Table \ref{tab:RMSE_NEMO}. In Figure \ref{fig:NEMO_MOC} we show 20-year mean meridional overturning circulation (MOC) streamfunction \cite{cabanes2008mechanisms}. Both backscatter parameterizations improve the streamfunction near the latitude of WBC separation ($\sim 30^o N$), but Reynolds parameterization is also better in improving the northern circulation cell ($\sim 45^o N$). Additionally, we show that both backscatter parameterizations significantly improve the resolved eddy meridional heat flux near the latitude $\sim 30^o N$ (see Figure \ref{fig:NEMO_MOC}).

\clearpage

\section{Conclusions}
In this work, we perform careful a priori analysis of energy and enstrophy fluxes in 2D decaying turbulence and develop mixed subgrid parameterizations based on previous studies \cite{germano1986proposal, germano1991dynamic, vreman1994formulation, horiuti1997new}, but in the context of 2D fluids.
%propose new subgrid parameterizations which are consistent with the physics of the 2D turbulent flow. 
We evaluate these parameterizations in a posteriori experiments for a range of resolutions and implement the Reynolds part of the new parameterization to quasi-geostrophic and primitive equation ocean models. 

Our main contributions and findings are as follows:
\begin{itemize}
    \item We consider the budget of subgrid KE \cite{jansen2014parameterizing} and estimation of subgrid KE \cite{khanigradient} to predict the domain-averaged kinetic energy flux produced by subgrid eddies, which is required to propose backscatter parameterization.
    %Methods of estimation of the subgrid kinetic energy can be successfully used to predict the domain-averaged energy exchange with subgrid eddies.
    \item The components of Germano decomposition play a special role in forming energy and enstrophy subgrid fluxes: Leonard and Cross stresses are responsible for the enstrophy dissipation; all three stresses (Leonard, Cross, Reynolds) contribute to the kinetic energy backscatter in large scales, but only the Reynolds stress produces almost positive-definite kinetic energy transfer.
    \item We start from the dynamic Smagorinsky model (DSM) in a priori analysis and show by gradual changes how to build a subgrid model which correctly simulates energy and enstrophy fluxes. In particular, we simulate the enstrophy dissipation by the Leonard stress and the biharmonic Smagorinsky model which approximates the Cross stress; an approximation to the Reynolds stress is used to simulate a missing backscatter of kinetic energy.
    \item The new subgrid parameterization (DMM+Reynolds) is numerically stable at zero molecular viscosity. It improves the reproduction of the kinetic energy spectrum and decay of enstrophy. The new method to estimate the subgrid energy flux allows to reproduce the growth of the resolved kinetic energy at a very high Reynolds number.
    \item The role of the Reynolds stress model as a kinetic energy backscatter parameterization holds in two additional ocean models: pseudospectral QG model and finite volume primitive equation model NEMO. Similarly to the \citeA{jansen2014parameterizing} backscatter parameterization, the Reynolds model allows to energize the flow and improves various statistical properties, such as KE spectrum, the vertical profile of EKE, interscale KE and APE transfers, the resolved meridional eddy heat flux, MOC and errors in surface fields, such as SST, SSS and SSH. 
\end{itemize}

The important result of our analysis is that our subgrid parameterizations do not contain free physical parameters. The only parameter that was tuned a posteriori is the filter-to-grid width ratio (FGR) which was shown to control the relative importance of the numerical discretization errors. We expect that the dynamic procedure to determine the Smagorinsky coefficient ($C_S$) can be extended to more complex governing equations. Various approaches can be proposed to choose backscattering coefficient ($C_R$): the procedure here suggested (Eqs. \eqref{eq:subgrid_KE_equation} and \eqref{eq:khani_model}) can be extended with realistic sinks and sources of subgrid KE \cite{jansen2019toward}; $C_R$ can be determined dynamically \cite{horiuti1997new, yuan2020deconvolutional} or can be used as a tunable parameter. Importantly, $C_R$ can be easily tuned manually, and in our experiments with the NEMO ocean model we showed it can be chosen uniquely for all depths and spatial locations.

%Further studies of Germano decomposition can gain new insights. 
Additional future studies related to Germano decomposition can allow us to gain new insights on subgrid modeling.
For example, we have shown that the role of the Reynolds stress model increases as the filter gets wider, as it is seen in the a priori MSE metric, simulation of resolved kinetic energy a posteriori, and better results in the northern region in NEMO ocean model, where Rossby deformation radius falls within subgrid scales. Consequently, new subgrid parameterizations for coarse models can be based on the prediction of the Reynolds stress instead of the full subgrid forcing, see for example, in the context of machine learning \cite{bolton2019applications, zanna2020data}. We demonstrated that the most severe discrepancy between predicted and diagnosed Smagorinsky coefficient comes not from the lack of scale invariance, but from a difference between a priori and a posteriori performance of the same dynamic model. We suggest that the crudest approximation in our DMM+Reynolds model comes from the representation of the Cross stress. New accurate models of the Cross stress could potentially improve the consistency of a priori and a posteriori experiments, and gain a posteriori performance.

\clearpage

\appendix 

\section{Estimation of subgrid energy flux} \label{appendix:energy_flux}
We decompose velocity gradient tensor $\frac{\partial \overline{u}_i}{\partial x_j}$
into symmetric $\overline{S}_{ij}$ and antisymmetric $\overline{\Omega}_{ij}$ parts, and consequently $\frac{\partial \overline{u}_i}{\partial x_j} \frac{\partial \overline{u}_i}{\partial x_j} = \overline{S}_{ij}\overline{S}_{ij}+\overline{\Omega}_{ij}\overline{\Omega}_{ij} = |\overline{S}|^2/2+\overline{\omega}^2/2$ \cite{borue1998local}. On a periodic domain, we have $\langle |\overline{S}|^2 \rangle = \langle \overline{\omega}^2 \rangle$ \cite{buxton2011interaction}.  Finally, the estimation of subgrid KE (Eq. \eqref{eq:khani_model}) is related to the resolved enstrophy ($Z = \overline{\omega}^2/2$) as:
\begin{equation}
	\langle e \rangle = \frac{\overline{\Delta}^2}{12} \langle Z \rangle. \label{SFS_energy_estimation}
\end{equation}

At a high Reynolds number, the resolved enstrophy can be lost only to the subgrid eddies, and thus $\frac{d}{d t} \langle Z \rangle = - \langle \Pi_Z \rangle$. Combining it with 
$\frac{d}{d t} \langle e \rangle =  \langle \Pi_E \rangle$ (Eq. \eqref{eq:subgrid_KE_equation}) and \eqref{SFS_energy_estimation}, we obtain:
\begin{equation}
	\langle \Pi_E \rangle = \frac{d}{d t} \langle e \rangle = \frac{\overline{\Delta}^2}{12} \cdot \frac{d}{d t} \langle Z \rangle =  - \frac{ \overline{\Delta}^2}{12} \langle \Pi_Z \rangle.
\end{equation}

\section{Eliminating numerical errors with explicit filtering approach} \label{sec:explicit_filtering}
\begin{figure}
\centerline{\includegraphics[width=0.6\textwidth]{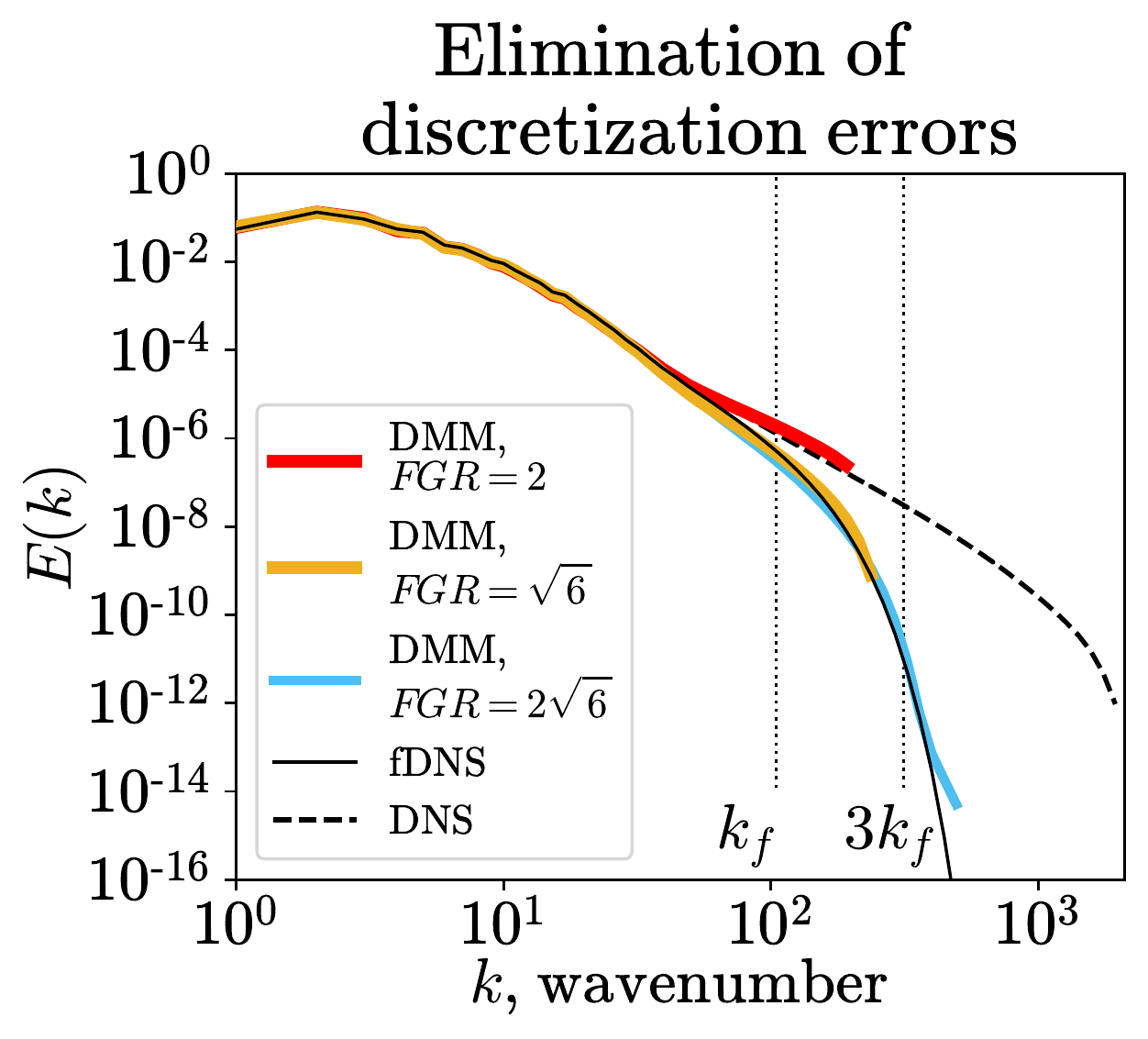}}
	\caption{Eliminating discretization errors in a posteriori experiments by enlarging the $FGR=\overline{\Delta}/\Delta_g$ parameter, where filter width is fixed ($\overline{\Delta}=\Delta_{512}$) and grid step is varying; $t=10$. Experiments with FGR equal to $2$, $\sqrt{6}$ and $2\sqrt{6}$ have resolutions $418^2$, $512^2$, $1024^2$, respectively.} 
 % The subgrid model DMM attempts to reproduce the filtered solution (fDNS) for narrow filter $\Delta_{512}$ and $t=10$. 
	\label{grid_convergence}
\end{figure}
%In Section \ref{LES_models} we have shown that the proposed subgrid models are accurate in a priori analysis. However, a priori metrics poorly correlate with a posteriori results \cite{ross2023benchmarking}. We suggest that the
The discretization errors may be an important source of discrepancies between a priori and a posteriori performance. As an example of numerical effects we refer to works of \citeA{bachman2017scale, maulik2017stable, guan2021stable}, where the accumulation of energy near the grid scale is shown for dynamic models. In this section, we apply the explicit filtering approach to reduce the role of numerical errors \cite{gullbrand2003effect}. The main idea of explicit filtering consists in considering the grid step of the coarse model $\Delta_g$ and filter width $\overline{\Delta}$ as independent parameters. Fixing the filter width $\overline{\Delta}$ and enlarging the $FGR=\overline{\Delta}/\Delta_g$, it is possible to eliminate the discretization errors from the LES equation \eqref{LES_eq}.

In Figure \ref{grid_convergence} we show the energy spectrum in a posteriori experiments at a fixed filter width and enlarging FGR (and corresponding grid resolution). At low resolution ($FGR=2$) we observe a build-up of energy density near the grid scale, and at larger FGRs coarse models converge to the filtered solution. There is a tradeoff between the strength of discretization errors and the number of directly simulated degrees of freedom \cite{lund1997use, lund2003use, bose2010grid, sarwar2017linking}. We use as small FGR as possible to better utilize the grid resolution, but large enough to reduce the role of discretization errors: a suitable choice is $FGR=\sqrt{6}$. It is also the maximum allowable FGR to use 3-point discrete filter defined in Eq. \eqref{discrete_filter}. We note that the optimal FGR depends on the hydrodynamic solver and can change in other configurations. The optimal FGR for mixed models (DMM and DMM+Reynolds) at all filter widths is $FGR=\sqrt{6}$, and DSM can be used with $FGR=2$. For convenience, we use $FGR=\sqrt{6}$ everywhere.

\section*{Data Availability Statement}
The software of the barotropic model in C++, QG model in Python and NEMO model in Fortran with implemented parameterizations are available via Zenodo \cite{perezhogin_dataset}, where we also provide simulation data and Figure plotting functions.

\acknowledgments
An analysis of high-resolution simulations (Sections 2 and 3) was carried out with the financial support of the Russian Science Foundation (project 21-71-30023). Development of subgrid closures (Sections 4 and 5)	was supported by the Moscow Center of Fundamental and Applied Mathematics at INM RAS (with the Ministry of Education and Science of the Russian Federation, Agreement 075-15-2022-286). %Implementation to realistic ocean models (Section 6) was supported by the generosity of Eric and Wendy Schmidt by recommendation of Schmidt Futures, as part of its Virtual Earth System Research Institute (VESRI).

We are grateful to Laure Zanna, Fabrizio Falasca, Gordey Goyman, Abigail Bodner, Elizabeth Yankovsky, Adam Subel and Dhruv Balwada for insightful discussions and help with the manuscript, and to Andrew Ross for help with implementation into the pyqg model. 

\bibliography{agusample}
\end{document}